\documentclass[useAMS,usenatbib,fleqn]{mnras}
\usepackage{mathtools,amssymb,amsmath,amsopn,graphicx,epsfig,subfigure, psfrag, natbib}
\usepackage{amsopn,epsfig,subfigure, psfrag, natbib}
\newcommand{\HI}{{{\rm HI~}}}
\title[Simulating the 21-cm visibility signal]{An analytical method to simulate the \HI $21$-cm visibility
  signal for intensity mapping experiments}

\author[Sarkar, Bharadwaj and Marthi]  {Anjan Kumar
  Sarkar$^{1}$\thanks{E-mail:anjan@cts.iitkgp.ernet.in}, Somnath
  Bharadwaj$^{2}$\thanks{E-mail:somnath@phy.iitkgp.ernet.in} and
  Visweshwar Ram Marthi$^{3,4,5}$\thanks{E-mail:vrmarthi@cita.utoronto.ca}
  \\ $^{1}$Centre for Theoretical Physics, IIT
  Kharagpur 721 302, India\\ $^{2}$Department of Physics \& Centre for Theoretical
  Physics, IIT Kharagpur 721 302, India\\ $^{3}$National Centre for Radio
  Astrophysics, Tata Institute of Fundamental Research, Pune 411 007,
  India\\ $^{4}$Canadian Institute for Theoretical Astrophysics, 60 St. George
  Street, Toronto, ON M5S 3H8, Canada\\ $^{5}$Dunlap Institute for Astronomy \&
  Astrophysics, 50 St. George Street, Toronto, ON M5S 3H4, Canada  }
\date{Accepted XXX. Received YYY; in original form ZZZ}

\pubyear{2017}

\begin{document}
\label{firstpage}
\pagerange{\pageref{firstpage}--\pageref{lastpage}}
\maketitle

\begin{abstract}
Simulations play a vital role in testing and validating \ion{H}{I} 21-cm
 power spectrum estimation techniques. Conventional methods
   use techniques like N-body simulations to simulate the sky signal
   which is then passed through a model of the instrument. This
 makes it necessary to simulate the \ion{H}{I} distribution in a large
 cosmological volume, and incorporate both the light-cone effect and
 the telescope's chromatic response. The computational requirements
 may be particularly large if one wishes to simulate many realizations
 of the signal. In this paper we present an analytical method to
 simulate the \ion{H}{I} visibility signal.  This is particularly efficient if
 one wishes to simulate a large number of realizations of the
 signal. Our method is based on theoretical predictions of the
 visibility correlation which incorporate both the light-cone effect
 and the telescope's chromatic response.  We have demonstrated this
 method by applying it to simulate the \ion{H}{I} visibility signal for the
 upcoming Ooty Wide Field Array Phase I.
 \end{abstract}





\section{Introduction}
The neutral hydrogen (\ion{H}{I}) in the diffuse intergalactic medium (IGM)
is ionized at redshifts $z \leq 6$ which is referred to as the
post-reionization era in the history of the universe. The residual \ion{H}{I}
is locked within dense pockets which are identified as the Damped
Lyman-$\alpha$ systems (DLAs) in quasar observations and have \ion{H}{I}
column densities ${\rm N_{\ion{H}{I}}} \geq 2 \times 10^{20} \, {\rm
  atoms/cm^{2}}$ \citep{wolfe2005,lanzetta1995,storrie1996}. \ion{H}{I} galaxy
surveys \citep{zwaan2005,martin2010}, DLA observations
\citep{rao2006,meiring2011} and \ion{H}{I} stacking
\citep{lah2007,delhaize2013,rhee2013,rhee2016} supply measurements of
$\Omega_{\rm g}$ at low redshifts ($z \leq 1$), while at high redshifts
($1 < z <6$) these measurements come from DLA
observations \citep{prochaska2009,noterdaeme2012,zafar2013}. These
measurements suggest $\Omega_{{\rm g}} \sim 10^{-3}$ to be almost
constant over the redshift range $1 < z < 6$, which implies a neutral
hydrogen fraction $\bar{x}_{{\rm HI}} = 0.02$.

Unlike traditional galaxy redshift surveys, \ion{H}{I} 21-cm measurements
cannot resolve individual \ion{H}{I} galaxies at higher redshifts due to the
limited angular resolution and sensitivity of present day radio
telescopes. The collective emission from the \ion{H}{I} sources appears as a
diffuse background radiation in low frequency radio observations below
$1420 \, {\rm MHz}$. The angular and frequency domain fluctuations in
this background radiation which are, in general, quantified through
the \ion{H}{I} 21-cm power spectrum have the potential to probe the large
scale structure of the universe at high $z$
\citep{bharadwaj2001,bharadwaj-sethi2001,wyithe-loeb2008}.  This
  technique, widely known as 21-cm intensity mapping (IM), makes it
  possible to use low frequency radio telescopes to survey larger
  cosmological volumes (e.g.
  \citealt{mcquinn2006,seo2010,ansari2012,bull-ferreira2015}).

Measurements of the post-reionization \ion{H}{I} 21-cm power spectrum hold the
prospect of measuring the baryon acoustic oscillation (BAO), which
can be used to constrain models of dark energy
\citep{wyithe2008,chang2008,seo2010,ansari2011}.  Measurement of the
\ion{H}{I} power spectrum can also be used to constrain the background
cosmological model without reference to the measurements of the BAO
\citep{bharadwaj2009,visbal2009}. The \ion{H}{I} 21-cm power spectrum
measurement can further be used to place constraints on the neutrino
mass \citep{loeb-wyithe2008,navarro2015,bharadwaj-sarkar2016}.
\citet{switzer2013} have claimed a detection of the \ion{H}{I} 21-cm intensity
fluctuations from $z \sim 0.8$ using single dish observations with the
Green Bank Telescope (GBT). Apart from the \ion{H}{I} 21-cm power spectrum, higher order statistics
such as the 21-cm bispectrum hold the prospect of quantifying the
non-Gaussianities in the \ion{H}{I} signal \citep{ali-pandey2005,hazra2012}.

A number of radio-interferometric arrays like
the Canadian Hydrogen Intensity Mapping Experiment \citep[CHIME\footnote{http://chime.phas.ubc.ca/};][]{bandura2014},
Baryon Acoustic Oscillation Broadband and Broadbeam array \citep[BAOBAB\footnote{http://bao.berkeley.edu/};][]{pober2013}, the
Tianlai project\footnote{http://tianlai.bao.ac.cn/}
\citep{chen2012,zhang2016} and 
  Hydrogen Intensity and Real-time Analysis eXperiment
  \citep[HIRAX\footnote{http://www.acru.ukzn.ac.za/~hirax};][]{newburgh2016} are
  being planned to measure the BAO using \ion{H}{I}
21-cm measurements in the redshift range $0.5 \leq z \leq 2.5$. Single
dish antennas are also being considered
\citep[BINGO\footnote{http://www.jb.man.ac.uk/research/BINGO/};][]{battye2012,battye2016}
to measure the BAO using 21-cm IM observations. The
Giant Metrewave Radio Telescope \citep[GMRT\footnote{http://gmrt.ncra.tifr.res.in/};][]{Swarup1991} is
capable of functioning at several different frequencies from $150 \,
{\rm MHz}$ to $1420 \, {\rm MHz}$, which correspond to measuring the
\ion{H}{I} radiation from the redshift range $0 \leq z \leq 8.5$. The
prospects of detecting the \ion{H}{I} signal using the GMRT have been studied
in earlier works \citep{bharadwaj2003,bharadwaj-ali2005}. Preliminary
observations using GMRT have been reported in
\citet{Ghosh2011a,ghosh2011b}. The GMRT is currently being upgraded to
operate at a larger bandwidth, forecasts for measuring the \ion{H}{I} signal
using the upgraded GMRT (uGMRT) are presented in Chatterjee et al. (in
preparation). LOFAR\footnote{http://www.lofar.org/}
  \citep{lofar2013}, MWA\footnote{http://www.mwatelescope.org/}
  \citep{mwa2013},
  PAPER\footnote{http://eor.berkeley.edu/}\citep[][]{parsons2010,aguirre2014}
  and the upcoming HERA\footnote{http://reionization.org/}
  \citep{deboer2017} all aim to carry out 21-cm IM observations to
  measure the Epoch of Reionization signal. Future radio telescopes
  like the
  SKA1-mid\footnote{https://www.skatelescope.org/multimedia/image/ska-infographic-ska1-mid/}
  and the
  SKA1-low\footnote{https://www.skatelescope.org/multimedia/image/ska-infographic-ska1-low/}
  hold promising prospects for measuring the post-reionization \ion{H}{I}
  21-cm power spectrum at an unprecedented level of precision
  \citep{guhasarkar-datta2015,navarro-viel2015,bull2015,santos2015,navarro-viel2016}.
  IM experiments are also being planned with the CO line (COMAP,
  \citealt{comap16}) and the \ion{C}{II} line (TIME, \citealt{crites2014}).

The Ooty Radio Telescope (ORT) \citep{swarup1971,sarma1975} is
currently being upgraded to operate as a linear radio-interferometric
array the Ooty Wide Field Array (OWFA)
\citep{prasad2011a,prasad2011b,subrahmanya2017b}. The primary science
goals of OWFA have been outlined in \citet{subrahmanya2017a}, and the
measurement of the $z=3.35$ post-reionization \ion{H}{I} 21-cm power spectrum
is one of its major objectives. It has been predicted
\citep{bharadwaj-fisher2015} that a $5 \sigma$ detection of the
amplitude of the \ion{H}{I} 21-cm power spectrum is possible with $\sim 150 \,
{\rm hrs}$ of observation. Further, \citet{sarkar-fisher2017} predict
that a $\sim 5 \sigma$ measurement of the binned \ion{H}{I} 21-cm power
spectrum is possible in the $k$-range $0.05 \leq k \leq 0.3 \, {\rm
  Mpc}^{-1}$ with $1,000 \, {\rm hrs}$ of observation.

The complex visibilities are the primary quantities which are directly
measured in radio-interferometric observations. The measured
visibilities have contributions from the \ion{H}{I} signal, foregrounds and
system noise. It is important and worthwhile to quantify the \ion{H}{I} signal
in terms of their expected contribution to the measured visibilities
(e.g. \citealt{bharadwaj2003,bharadwaj-srikant2004}) Detailed
predictions for the visibility correlations at different baselines and
frequency channels for the expected statistical \ion{H}{I} signal, foregrounds
and the system noise expected at OWFA are presented in
\citet{ali-bharadwaj2014} and \citet{gehlot-bagla2017}.  Theoretical
estimates presented in these papers, and also direct observations
(e.g. \citealt{Ghosh2011a,ghosh2011b}) predict the foreground
contribution to be several orders of magnitude larger than the \ion{H}{I}
signal at OWFA.  It is therefore crucial to consider foreground
removal for the \ion{H}{I} 21-cm signal. The various astrophysical
  foregrounds are expected to have a smooth frequency dependence. In
  contrast to this the expected \ion{H}{I} visibility signal at two different
  frequencies decorrelates rapidly as the frequency separation is
  increased, and the visibility correlation is predicted to decay to a
  value very close to zero within a frequency separation of $\sim 4 \,
  {\rm MHz}$ at OWFA \citep{ali-bharadwaj2014}.  This is a generic
  feature of the 21-cm signal
  \citep{bharadwaj-sethi2001,bharadwaj-ali2005} and most of the
foreground removal techniques
\citep[e.g.][]{mcquinn2006,jelic2008,gleser2008,bowman2009,parsons2012}
use this feature to distinguish between the foregrounds and the \ion{H}{I}
signal. However, the chromatic response of the telescope introduces
frequency-dependent structures in the foregrounds, and it is important
to model these in order to quantify these effects in any foreground
removal technique. 
Simulations incorporating the expected foregrounds, \ion{H}{I} signal and
various instrumental and post-processing effects are crucial in
validating any measurement technique for the \ion{H}{I} 21-cm power spectrum.
\citet{marthi2017} presents a software model for OWFA which has been
developed with a view to simulate the visibilities, as well as processing of the
visibility data including calibration\citep[see][for the calibration
  algorithm]{marthi-chengalur2014} and power spectrum estimation. Simulated foreground predictions for OWFA are presented
in \citet{marthi-chatterjee2017} where, in addition, they present a detailed
analysis of the instrument-induced systematics and discuss the multi-frequency
angular power spectrum estimator (MAPS) for OWFA. \citet{chatterjee2017} present
simulations of the \ion{H}{I} 21-cm visibility signal expected for OWFA.

Numerical simulations of the visibilities would typically start with a
simulation of the expected sky signal which would then be passed
through a software model of the instrument.  While this is relatively
straightforward for the foregrounds, it may be computationally challenging for
the redshifted 21-cm signal where it is necessary to simulate the \ion{H}{I}
distribution in a large cosmological volume. The fact that the \ion{H}{I}
signal and the cosmological parameters both evolve along the line of
sight due to the light-cone effect introduces further computational
complications. The difficulty is compounded when it is necessary to simulate
several statistically independent realizations of the signal, which an analytical
prescription could substantially reduce.


In the present work we provide a prescription for
simulating the \ion{H}{I} signal visibilities and validate the method for
OWFA. Our method, which is described in Section 2 of this paper, uses
the analytically predicted visibility correlation as an input. In
addition to the chromatic response of the telescope, we 
incorporate the redshift evolution of both the \ion{H}{I} signal and the
cosmological parameters within the observing frequency bandwidth.  The
visibility correlation matrix is decomposed into its eigenbasis or the
Koshambi-Karhunen-Lo{\'e}ve (KKL)-basis \citep{turin2011} which is
then used to simulate different random realizations of the \ion{H}{I}
visibilities. Section 3 gives a brief overview
of OWFA and Section 4 presents the \ion{H}{I} model used in our analysis.
Section 5 presents the results, and we present the summary and
discussion in Section 6. We use the \citet{eisenstein-hu1998} transfer function to calculate
 the cosmological matter power spectrum and adopt the \emph{PLANCK+WMAP9}
 best-fit cosmological parameters from \citet{ade2014}.

\section{Simulation of the \HI visibilities: Analytical Method}
The visibilities $ \mathbfit{V}(\mathbfit{U}, \nu)$ measured at different
baselines $\mathbfit{U}$ and frequency channels $\nu$ may be considered to be
the sum of three different components
\begin{equation}
\mathbfit{V} = \mathbfit{S} + \mathbfit{F} + \mathcal{N}
\end{equation}
namely, the \ion{H}{I} signal $\mathbfit{S}$, foregrounds $\mathbfit{F}$ and
noise $\mathcal{N}$ respectively. The noise in each visibility
measurement may be modelled as an independent Gaussian random variable
with zero mean and whose variance is known
\citep{book-GMRT,thompson2008}.  The foregrounds, which largely
originate from our Galaxy and external radio galaxies, are found to be
$4$ to $5$ orders of magnitude larger than the \ion{H}{I} signal
\citep{ghosh2012}. In this paper we focus on simulating only the \ion{H}{I}
signal $\mathbfit{S}$, and we do not consider the two other components.

The baseline $\mathbfit{U} =\mathbfit{d}/\lambda$  for a fixed antenna spacing
$\mathbfit{d}$ varies with the observing frequency $\nu$. This poses a problem
in the notation if we attempt to interpret the visibility $
\mathbfit{V}(\mathbfit{U}, \nu)$ as a function of two independent variables $\mathbfit{U}$
and $\nu$. We avoid this problem by holding $\mathbfit{U}$ fixed at the value
calculated at the central frequency $\mathbfit{U} =\mathbfit{d}/\lambda_c$ and
expressing $\mathbfit{d}/\lambda$ as $\mathbfit{U} \nu/\nu_c$. The \ion{H}{I} signal contribution to the visibility can then be
expressed as
\begin{equation}
\mathbfit{S}(\mathbfit{U}, \nu) = \left( \frac{\partial B}{\partial T}
\right)_{\nu} \, \int d^2 \theta \, A(\boldsymbol{\theta}, \nu) \, \delta T_b(\boldsymbol{\theta},\nu)
\, e^{2 \pi i \boldsymbol{\theta} \cdot \mathbfit{U} \nu/\nu_c}
\label{eq:2}
\end{equation}
where $\left(\frac{\partial B}{\partial T} \right)_{\nu} = 2 k_B
\nu^2/c^2$ is the conversion factor from temperature to specific
intensity, $A(\boldsymbol{\theta}, \nu)$ is the the primary beam pattern of the
antennas and $\delta T_b(\boldsymbol{\theta},\nu)$ is the redshifted \ion{H}{I} 21-cm
brightness temperature fluctuations which trace the cosmological \ion{H}{I}
distribution at the redshift $z$ where the 21-cm radiation originated.
$\delta T_b(\boldsymbol{\theta},\nu)$ is usually modelled as a random field which is
quantified through various statistical quantities like the two-point
correlation function or equivalently the power spectrum. It is
also clear that we expect the different visibilities $\mathbfit{S}(\mathbfit{U},
\nu)$ to be random variables whose statistical properties reflect
those of the underlying \ion{H}{I} distribution.

The aim of this paper is to present a method to simulate
$\mathbfit{S}(\mathbfit{U}, \nu)$. To this end, we predict the statistical
properties of $\mathbfit{S}(\mathbfit{U}, \nu)$ and subsequently use these
predictions to simulate $\mathbfit{S}(\mathbfit{U}, \nu)$.  Below, we first
illustrate the method by applying it to a simplified situation and we
subsequently consider a more general situation.

\subsection{A simplified analysis}
\label{ss1}
 We focus our attention on a fixed baseline $\mathbfit{U}$ and consider
the visibilities $\mathbfit{S}(\mathbfit{U}, \nu)$ at different
frequencies.  We make two assumptions which introduce
substantial simplifications. First, we assume that the frequency
dependence of the visibility $\mathbfit{S}(\mathbfit{U}, \nu)$ arises entirely
from the 21-cm brightness temperature fluctuations $\delta
T_b(\boldsymbol{\theta},\nu)$ and we ignore the frequency dependence of all the other
terms in the right hand side of ~(\ref{eq:2}) whereby
\begin{equation}
\mathbfit{S}(\mathbfit{U}, \nu) = \left( \frac{\partial B}{\partial T}
\right)_{\nu_c} \, \int d^2 \theta \, A(\boldsymbol{\theta}, \nu_c) \, \delta
T_b(\boldsymbol{\theta},\nu) \, e^{2 \pi i \boldsymbol{\theta} \cdot \mathbfit{U}} \,.
\label{eq:2a}
\end{equation}
This assumption essentially ignores the chromatic response of the
telescope. Second, we model $\delta T_b(\boldsymbol{\theta},\nu) \equiv \delta
T_b(\mathbfit{x})$ as a statistically homogeneous Gaussian random field. The
statistical homogeneity is with respect to $\mathbfit{x} \equiv
(\mathbfit{x}_{\perp},x_{\parallel})= (r_{\nu_c} \boldsymbol{\theta}, r^{'}_{\nu_c}
(\nu-\nu_c))$ which denotes the comoving separation between
$(\boldsymbol{\theta},\nu)$ and the center of the observational volume which is
located at $(\boldsymbol{\theta}=0,\nu_c)$.  Here we use $r_{\nu}$ to denote the
comoving distances corresponding to the redshifted 21cm radiation
received at the frequency $\nu$, and $r_{\nu_c}$ and $r^{'}_{\nu_c}$
respectively refer to $r_\nu$ and $r^{\prime}_{\nu}=d r_\nu/d \nu$ evaluated at
$\nu=\nu_c$. The assumption of statistical homogeneity implies that
the different Fourier modes $\tilde{\Delta} T_b(\mathbfit{k})$ and
$\tilde{\Delta} T_b^{*}(\mathbfit{k}^{'})$ are uncorrelated
\citep{peebles1980}. The
statistical properties of $\delta T_b(\mathbfit{x})$ can be completely
quantified using the 3D 21-cm
brightness temperature power spectrum $P(\mathbfit{k})$, defined
through
\begin{equation}
\langle \tilde{\Delta} T_b(\mathbfit{k}) \tilde{\Delta} T_b^{*}(\mathbfit{k}^{'}) \rangle
= (2 \pi)^3 \, P(\mathbfit{k}) \, \delta_D^3(\mathbfit{k} - \mathbfit{k}^{'})
\end{equation} 
where $\delta_D^3(\mathbfit{k} - \mathbfit{k}^{'})$ is the 3D Dirac delta function. It may
be noted that the different Fourier modes will, in general, be
correlated if the assumption of statistical homogeneity is broken.

We use the two visibility correlation
\begin{equation}
\mathbfss{S}_2(\mathbfit{U},\nu_a,\nu_b)=\langle
\mathbfit{S}(\mathbfit{U}, \nu_a) \mathbfit{S}^{*}(\mathbfit{U}, 
\nu_b) \rangle
\label{eq:3a}
\end{equation}
to quantify the statistical properties of the visibilities. The two
assumptions adopted earlier imply that the visibilities
$\mathbfit{S}(\mathbfit{U}, \nu)$ are statistically homogeneous in frequency,
whereby $\mathbfss{S}_2(\mathbfit{U},\nu_a,\nu_b)$ depends only on the frequency separation
$\nu_a-\nu_b$.  Calculating $\mathbfss{S}_2(\mathbfit{U},\nu_a,\nu_b)$ in terms of the
power spectrum $P(\mathbfit{k})$ \citep{bharadwaj-ali2005}, we have
\begin{flalign}
\mathbfss{S}_2(\!\mathbfit{U}, \nu_a, \nu_b\!) = \left(\!\frac{\partial B}{\partial T}
\! \right)^2_{\nu_c}\! \int\!\frac{d^3k}{(2\pi)^3} \left| \tilde{A}\! \left(\!\mathbfit{U} -
\frac{\mathbfit{k}_{\perp} r_{\nu_c}}{2\pi},\nu_c\! \right) \right|^2_{\nu_c} \nonumber &
\\
P(\mathbfit{k}) \, e^{i r_c^{'} k_{\parallel} (\nu_b - \nu_a)},
\label{eq:3}
\end{flalign}
where $\tilde{A}(\mathbfit{L}, \nu) $ is the Fourier transform of the primary beam
pattern $A(\boldsymbol{\theta},\nu)$ with $\mathbfit{L}$ being the Fourier conjugate to $\boldsymbol{\theta}$,
and $\mathbfit{k}_{\perp}$ and $k_{\parallel}$ are respectively the components
of the wave vector $\mathbfit{k}$ perpendicular and parallel to the line of
sight. We use ~(\ref{eq:3}) to predict the statistical properties
of the visibilities.  The properties of the visibility correlations
have been studied in several other works \citep{bharadwaj2003,bharadwaj-ali2005,ali-bharadwaj2014}.
We note that since the visibilities at different frequencies are correlated, it is not
possible to independently simulate them.

It is convenient to decompose the visibilities into delay channels
\citep{Morales2005}
\begin{equation}
\mathbfit{S}(\mathbfit{U}, \nu) = \int \, \mathbfit{s}(\mathbfit{U}, \tau) \, e^{-2 \pi i \tau \nu} \,
d\tau \,.
\label{eq:4}
\end{equation}
The statistical homogeneity of $\mathbfit{S}(\mathbfit{U}, \nu)$ along frequency
implies that the different delay (or Fourier) modes $\mathbfit{s}(\mathbfit{U}, \tau)$ are
uncorrelated
\begin{equation}
\langle s(\mathbfit{U},\tau) \mathbfit{s}^{*}(\mathbfit{U},\tau^{'}) \rangle =
\delta_D(\tau-\tau^{'}) p(\mathbfit{U},\tau)
\label{eq:5}
\end{equation}
where $p(\mathbfit{U},\tau)$ is the power spectrum of $\mathbfit{s}(\mathbfit{U},
\tau)$. Using ~(\ref{eq:4}) to calculate the visibility correlation
we have
\begin{equation}
\mathbfss{S}_2(\mathbfit{U}, \nu_a, \nu_b) = \int p(\mathbfit{U},\tau) \, e^{2 \pi i \tau
  (\nu_b-\nu_a)} \, d \tau \,.
\label{eq:6}
\end{equation}

Comparing ~(\ref{eq:3}) with ~(\ref{eq:6}), it is possible to identify
$\tau= k_{\parallel} r^{'}_{\nu_c}/(2 \pi)$ and
\begin{equation}
p(\mathbfit{U},\tau)=\left(\! \frac{\partial B}{\partial T}\! \right)^2\! \frac{1}{r^{'}_{\nu_c}}\int\! \frac{d^2k_{\perp}}{(2\pi)^2} \left| \tilde{A}\!
\left(\!\mathbfit{U}\! -\! \frac{\mathbfit{k}_{\perp} r_{\nu_c}}{2\pi}, \nu_c\!\right)
\right|^2_{\nu_c}\! P(\mathbfit{k}).
\label{eq:7}
\end{equation}
The different $s(\mathbfit{U},\tau)$ are independent, and have amplitude
$\sqrt{p(\mathbfit{U},\tau)}$.  It is straightforward to simulate $s(\mathbfit{U},\tau)$
and use these to simulate the visibilities
 \begin{equation}
\mathbfit{S}(\mathbfit{U}, \nu) = \int \, \sqrt{\frac{p(\mathbfit{U}, \tau)}{2}} \,
        [x(\tau) + i y(\tau)] \, e^{-2 \pi i \tau \nu} \, d\tau
\label{eq:8}
\end{equation}
where, $x(\tau)$ and $y(\tau)$ are two independent Gaussian random
fields with zero mean and unit variance, that is $\langle x(\tau)
x(\tau^{'}) \rangle = \delta_D(\tau- \tau^{'})$. 

In summary, given a telescope we can use the input 21-cm power
spectrum $P(\mathbfit{k})$ along with eqs.~(\ref{eq:7}) and
(\ref{eq:8}) to 
simulate the visibilities at different frequencies for a fixed
baseline.

\subsection{A generalized analysis}
\label{ss2}
A radio-interferometric array typically has many different baselines
$\mathbfit{U}_i$ and its observing band is split into frequency channels $\nu_n$, with the possibility that the
signal in several of the baselines and across frequency channels may be correlated
\citep{ali-bharadwaj2014}. It is therefore necessary to
simultaneously consider all the visibilities that will be observed by the
array. Here we use $\mathbfit{S}_a \equiv \mathbfit{S}(\mathbfit{U}_a,\nu_a)$ to denote the
different data elements where the index $a$ refers to a combination of
baseline and frequency channel $(\mathbfit{U}_a,\nu_a)$ and $a=1,2,...,N$ spans
the entire visibility data observed by the array.  We now use the
two-visibility correlation
\begin{equation}
\mathbfss{S}_{2ab}=\langle \mathbfit{S}_a \mathbfit{S}^{*}_b \rangle
\label{eq:9}
\end{equation}
to quantify the statistical properties of the visibility data.

The present analysis incorporates the chromatic response of the
telescope, and we use ~(\ref{eq:2}) to calculate the
visibilities. We also incorporate the Light Cone (LC) effect which
essentially implies that the comoving distance $r_\nu$ and the look back
time both vary with the observational frequency $\nu$. This modifies
the mapping from $(\boldsymbol{\theta},\nu)$ to $\mathbfit{x}$, and we now have
$(\mathbfit{x}_{\perp},x_{\parallel})=(r_{\nu} \boldsymbol{\theta}, r_{\nu}-r_{\nu_c})$ which is a
non-linear relation compared to the simple linear approximation
used in \ref{ss1},
breaking the
statistical homogeneity of $\delta T_b(\mathbfit{x})$ along the line of sight
direction $x_{\parallel}$. Here we model $\delta T_b(\boldsymbol{\theta},\nu)$ by
assuming that $\delta T_b(\mathbfit{x})$ is linearly related to $\delta(\mathbfit{x})$
which is the density contrast of the underlying matter distribution
normalized at the present epoch, the relation between $\delta T_b(\mathbfit{x})$
and $\delta(\mathbfit{x})$ containing factors which incorporate the cosmological
evolution and redshift space distortion. The fluctuations $\delta(\mathbfit{x})$
may be assumed to be statistically homogeneous and isotropic, and it
is convenient to consider their Fourier transform
$\Delta(\mathbfit{k})$. Statistical homogeneity implies that the different
Fourier modes $\Delta(\mathbfit{k})$ and $\Delta(\mathbfit{k}^{'})$ are uncorrelated, and
we use $P(k)$ to denote the power spectrum of the underlying matter
fluctuations $\Delta(\mathbfit{k})$.  Note that $P(k)$ is normalized at the
present epoch. This allows us to write
\begin{equation}
\delta T_b(\boldsymbol{\theta},\nu)=Q_{\nu} \int \frac{d^3 k}{(2 \pi)^3}
       [1+\beta_{\nu} \mu^2] \Delta(\mathbfit{k}) e^{-i [\mathbfit{k}_{\perp} \cdot \boldsymbol{\theta}
           r_{\nu} +k_{\parallel} (r_{\nu} -r_{\nu_c})]}
\label{eq:a1}
\end{equation}
where $Q_{\nu}=[\bar{T} b_{\rm HI} \bar{x}_{\rm HI} D]_{\nu}$ refers
to the product of a set of quantities namely $\bar{T}$ the mean 21-cm
brightness temperature \citep{bharadwaj-ali2005}, $b_{\rm HI}$ the
linear \ion{H}{I} bias, $\bar{x}_{\rm HI}$ the mean Hydrogen neutral fraction
and $D$ the growing mode of linear density perturbations, all of which
evolve with redshift $z$ or equivalently vary with the observational
frequency $\nu$.  The factor $[1+\beta_{\nu} \mu^2]$ incorporates the
effect of redshift space distortion \citep{kaiser1987}, here
$\mu=k_{\parallel}/k$ and $\beta_{\nu}$ is the linear redshift
distortion parameter which also varies with $\nu$.

We can now use ~(\ref{eq:2}) to calculate the
visibilities as
\begin{flalign}
\mathbfit{S}(\mathbfit{U},\nu) = Q_{\nu} \int \frac{d^3 k}{(2 \pi)^3}\, \tilde{A}\left(\mathbfit{U}
  \frac{\nu}{\nu_c} - \frac{\mathbfit{k}_{\perp} r_{\nu}}{2\pi},\nu \right)
  \nonumber &
  \\
[1+\beta_{\nu} \mu^2] \Delta(\mathbfit{k}) e^{-i k_{\parallel} (r_{\nu}
       -r_{\nu_c})}
\label{eq:a2}
\end{flalign}
and the visibility correlation as
\begin{flalign}
\mathbfss{S}_{2ab} = Q_{\nu_a} Q_{\nu_b} \int \frac{d^3 k}{(2 \pi)^3} \tilde{A}
\left(\mathbfit{U}_a \frac{\nu_a}{\nu_c} - \frac{\mathbfit{k}_{\perp}
  r_{\nu_a}}{2\pi},\nu_a \right) \nonumber &\\
\tilde{A}^* \left(\mathbfit{U}_b \frac{\nu_b}{\nu_c} -
\frac{\mathbfit{k}_{\perp} r_{\nu_b}}{2\pi},\nu_b \right) 
\times P_\mathrm{rsd}(k)  e^{-i k_{\parallel} (r_{\nu_b} -r_{\nu_a})},
\label{eq:a3}
\end{flalign}
where
\begin{equation}
 P_\mathrm{rsd}(k) = [1+\beta_{\nu_a} \mu^2] \, [1+\beta_{\nu_b} \mu^2] \, P(k)
\end{equation}
is the effective power spectrum considering the redshift space distortion.
The fluctuations $\delta(\mathbfit{x})$ are here assumed to be a Gaussian random
field, and the statistical properties of the visibilities are entirely
predicted by ~(\ref{eq:a3}).

It is quite obvious from ~(\ref{eq:a1}) that the visibilities
$\mathbfit{S}(\mathbfit{U},\nu)$ are not statistically homogeneous along frequency. In
this case, the Fourier transform can not be considered the ideal
basis. The Fourier basis used in \ref{ss1} also does not incorporate
the fact that the visibilities at different baselines may be
correlated. What then are the correct basis vectors which can be used
to decomposes the visibility signal into component which are
statistically independent? Kosambi \citep{Kosambi1943}, and
subsequently Karhunen \citep{karhunen1947} and Lo{\'e}ve
\citep{loeve1955}, showed that the correct basis is provided by the
eigenvectors $\hat{e}^{\alpha}_a$ of the data covariance matrix (the
visibility correlation)
\begin{equation}
\mathbfss{S}_{2ab} = \sum_{\alpha=1}^N \lambda_{\alpha}\, [\hat{e}^{\alpha}_a] \,
[{\hat{e}^{\alpha}_b}]^{\dagger}\,.
\label{eq:a4}
\end{equation}
where the index $\alpha$ refers to the different eigenvalues
  $\lambda_{\alpha}$ and $\alpha$ runs from $1$ to $N$ which is the
  total number of eigenvalues as well as the total number of
  visibility data.

It is convenient to decompose the visibility data $\mathbfit{S}_a$ in this ``KKL
basis'' $\hat{e}^{\alpha}_a$ using
\begin{equation}
\mathbfit{S}_a=\sum_{\alpha=1}^N s_{\alpha} \, \hat{e}^{\alpha}_a
\label{eq:a5}
\end{equation}
which is in exact analogy with ~(\ref{eq:4}). Here
  $s_{\alpha}$ are the components of the data vector $\mathbfit{S}_a$ in the
  $\hat{e}^{\alpha}_a$ basis. The different components $s_{\alpha}$
  are uncorrelated
\begin{equation}
\langle s_{\alpha} s^{*}_{\beta} \rangle =\delta_{\alpha,\beta}
\lambda_{\alpha} \,,
\label{eq:a6}
\end{equation}
and have amplitude $\sqrt{\lambda_{\alpha}}$, in exact analogy with
~(\ref{eq:6}).  Note that $\delta_{\alpha,\beta}$ here refers to
the Kronecker delta.
 
 It is now straightforward to simulate a random realization of the
  visibilities. We proceed by first generating a random realization of
  the $N$ components $s_{\alpha}$ using
\begin{equation}
s_{\alpha}=\sqrt{\frac{\lambda_\alpha}{2}} (x_{\alpha} + i y_{\alpha})
\,.
\label{eq:a7}
\end{equation}
Here $x_{\alpha}$ and $y_{\alpha}$ are $2N$ independent Gaussian
  random variables of unit variance $\langle x_{\alpha} x_{\beta}
  \rangle =\delta_{\alpha,\beta}$. These components $s_{\alpha}$
  are then used in ~(\ref{eq:a5}) to generate a random realization
  of the visibilities.  This is in exact analogy with ~(\ref{eq:8})
  which has been discussed earlier. The same process (\ref{eq:a7}
  and \ref{eq:a5}) can be repeated with different sets of the $2N$
  Gaussian random numbers $x_{\alpha}$ and $y_{\alpha}$ to simulate
  different random realizations of the visibility data
  $\mathbfit{S}_a$.

In summary, $\mathbfss{S}_{2ab}$ (\ref{eq:a3}) quantifies the statistical
properties of the entire visibility data. In order to simulate the
visibility data, it is necessary to numerically determine the
eigenvalues and eigenvectors of $\mathbfss{S}_{2ab}$. Random realizations of the
eigenvalues obtained through \ref{eq:a7} can hence be used to generate 
as many distinct realizations of the visibilities.

\section{The Ooty Wide Field Array (OWFA)}
Efforts are currently underway to upgrade the Ooty Radio Telescope
(ORT) into a linear radio-interferometric array
\citep{prasad2011a,prasad2011b,subrahmanya2017a,subrahmanya2017b}, the
Ooty Wide Field Array (OWFA). With a nominal frequency of $\nu_0 =
326.5 \, {\rm MHz}$, this directly corresponds to measuring the \ion{H}{I}
radiation from the redshift $z = 3.35$. The telescope, situated
  on the Nilgiri hills, is a parabolic cylindrical reflector of length
  $530 \, {\rm m}$ and width $30 \, {\rm m}$.  The long axis of the
  cylinder is tilted by $11^{\circ}$ which matches the telescope
  latitude of $11^{\circ}$ North. This makes the telescope's long axis
  parallel to the Earth's rotation axis \citep{swarup1971,sarma1975}
and enables the telescope to observe the same part of the sky with a
single rotation.  The telescope has a feed system that consists of
1056 half-wavelength dipoles which are placed almost end-to-end along
the length of the cylinder. The entire feed is placed off-axis to
avoid maximal obstruction to the incoming radiation.

An upgrade to the OWFA will allow operation in two concurrent
  interferometric modes - Phase I and Phase II
  \citep{prasad2011a,prasad2011b,ali-bharadwaj2014}. This study
  focuses on Phase I which has a total of $N_A = 40 $ antennas
  arranged linearly along the length of the cylindrical
  reflector. Each antenna is effectively a small parabolic cylindrical
  reflector of width $30 \, {\rm m}$ and length $11.5 \, {\rm m}$
  containing $24$ dipole elements along the focal line.  The antennas
  have a rectangular aperture of dimension $b \times d$, where $b = 30
  \, {\rm m} $ and $d=11.5 \, {\rm m}$.  We consider the telescope
aperture to lie on the $x-y$ plane with the $x$ axis along the length
of the cylinder. The aperture power pattern of OWFA can
approximately be written as\citep{ali-bharadwaj2014},
\begin{equation}
\tilde{A}(\mathbfit{L}, \nu) = \left( \frac{\lambda^2}{bd} \right) \, \Lambda\left(
\frac{L_x \lambda}{d} \right) \, \Lambda\left( \frac{L_y \lambda}{b}
\right)
\label{eq:b1}
\end{equation}
where $\lambda = c/\nu$ is the observing wavelength,
$\mathbfit{L}=(L_x,L_y)$ 
and the triangular function $\Lambda(w)$ has been defined in the usual
manner as, $\Lambda(w) = (1 - |w|)$ for $|w| < 1$ and zero elsewhere.
We note that aperture power pattern $\tilde{A}(\mathbfit{L}, \nu)$
peaks at $\mathbfit{L} = (0,0)$ 
and extends over the range $-{d}/{\lambda} \leq L_x \leq
{d}/{\lambda}$ along $L_x$ and the range $-{b}/{\lambda} \leq L_y \leq
{b}/{\lambda}$ along $L_y$, and is zero beyond, and has as such been
used to calculate the visibility covariance for OWFA.

OWFA is a linear interferometric array with all the antennas equally
spaced along the length of the ORT cylinder. This allows us to write
the OWFA baselines $\mathbfit{U}_n$ as
\begin{equation}
\mathbfit{U}_n = n \, \left( \frac{\mathbfit{d}}{\lambda_c} \right)
\label{eq:b2}
\end{equation}
where $\lambda_c$ is the central observing wavelength and $n$ is the
baseline index which can have values in the range $1 \leq n \leq
N_A-1$ where $N_A$ the total number of antennas in the array. The
smallest baseline measures $|\mathbfit{d}|=11.5$ m which is the spacing
between two adjacent antennas in the array. We have a total of
$(N_A-1)=39$ baselines for Phase I of OWFA.

Phase I has an operating bandwidth of $B=39 \, {\rm MHz}$. For the present
analysis we consider a smaller bandwidth $B=16$ MHz and  $N_c = 128$
frequency channels of width $\Delta \nu_c = 0.125 \, {\rm MHz}$ each to reduce
the computation time. Phase-I provides $4992$ (39 baselines $\times$ 128 channels)
instantaneous visibility measurements: the visibility correlation
matrix $\mathbfss{S}_{2ab}$ (\ref{eq:a3}) is hence Hermitian. The correlation matrix
\begin{equation}
\mathbfss{S}_{2ab}=\langle \mathbfit{S}(\mathbfit{U}_a,\nu_a) \mathbfit{S}^{*}(\mathbfit{U}_b,\nu_b) \rangle
\label{eq:c1}
\end{equation}
may be thought of as a combination of three different kinds of blocks.
We first consider $\mathbfss{S}_2(\mathbfit{U}_n,\nu_p,\nu_q)$ (\ref{eq:3}) which is
the correlation between visibilities measured at the same baseline
$\mathbfit{U}_a=\mathbfit{U}_b=u_n$ but the two frequencies $\nu_p$ and $\nu_q$ can
differ. There are $39$ such blocks each of which has $128 \times 128$
elements. Next, there are baseline pairs $\mathbfit{U}_a=\mathbfit{U}_m$ and $\mathbfit{U}_b=\mathbfit{U}_{n}$
with $m \neq n$. The correlation $\mathbfss{S}_{2ab}$ is non-zero if there is a
range of $\mathbfit{U}$ values where the aperture power patterns
$\tilde{a}(\mathbfit{U}_n-\mathbfit{U},\nu_p)$ and $\tilde{a}(\mathbfit{U}_m-\mathbfit{U},\nu_q)$ have an
overlap. For OWFA there is an overlap only for the adjacent baselines
$m=n \pm 1$ \citep[see Fig. 2 of][]{ali-bharadwaj2014}, and the
correlation
\begin{equation}
\mathbfss{S}_{2}(\mathbfit{U}_m,\nu_p,\mathbfit{U}_{n},\nu_q)=\langle \mathbfit{S}(\mathbfit{U}_m,\nu_p)
\mathbfit{S}^{*}(\mathbfit{U}_{n},\nu_q) \rangle
\label{eq:c2}
\end{equation}
has a non-zero value only when $m=n$ or $m=n \pm 1$.  The correlations
between the adjacent baselines are approximately one-fourth of those
between the same baselines\citep{bharadwaj-fisher2015} There are $38$
such blocks, each with $128 \times 128$ elements. The complex
conjugate of these blocks also appears in $\mathbfss{S}_{2ab}$.  The third kind
of blocks arise from visibility correlations beyond the adjacent
baselines.  The visibilities in these baselines are uncorrelated, and
all the elements of the blocks are zero.

\section{\HI model}
We now discuss the parameters of the input \ion{H}{I} model (\ref{eq:a1})
which has been used to compute the visibility covariance
(\ref{eq:a3}) for OWFA. Our \ion{H}{I} model is completely described by
the parameter $Q_{\nu} = [\bar{T} b_{\rm \HI} \bar{x}_{\rm \HI}
  D]_{\nu}$, the redshift distortion parameter $\beta$ and the matter
power spectrum $P(k)$. We have used the fit to the baryonic matter
power spectrum provided by \citet{eisenstein-hu1998} throughout our
analysis.  The bandwidth $B=16 \, {\rm MHz}$ used in our analysis
corresponds to a redshift range $3.24 \leq z \leq 3.46$. The
measurement $\Omega_{\rm \HI} = 10^{-3}$ from DLA observations
(e.g. \citealt{zafar2013}) combined with $\Omega_{b0} = 0.048$
\citep{ade2014} together imply $\bar{x}_{\HI} = 0.02$. Results from
simulations (e.g. \citealt{bharadwaj-sarkar2016}) and analytic
modelling \citep{marin2010} indicate a scale independent \ion{H}{I} bias
$b_{\rm \HI} = 2.0$ at large scales $k \leq 1 \, {\rm Mpc}^{-1}$.  We
assume that the values $\bar{x}_{\rm \HI} = 0.02$ and $b_{\rm \HI} = 2.0$ do
not evolve significantly over the relevant redshift range, and we have
held these fixed for our analysis. The values of $\bar{T}$ and $D(z)$
respectively vary across the range $17.41 \geq \bar{T} \geq 16.95$ and
$2.81 \times 10^{-1} \leq D(z) \leq 2.95 \times 10^{-1}$ over the
relevant redshift range, and the product $Q_{\nu}$ varies from $0.195$
to $0.200$.

The redshift distortion parameter is defined as $\beta = f(\Omega) /
b_{HI}$ where $f(\Omega)$ is the growth factor of the underlying
matter density fluctuations. Although accounting for the redshift evolution
of $f(\Omega)$, the resulting variation in our simulation is less than $0.5 \, \%$.

We note that the redshift evolution of $Q_{\nu}$ and $\beta$ is not
very significant in our analysis as the fractional bandwidth is quite small, in
contrast with observations which span a large bandwidth, e.g. an octave. However, the method presented
here correctly accounts for the predicted evolution of the \ion{H}{I} signal
and is expected to work well for such scenarios.

\section{Results}
\begin{figure*}
\begin{center}
\psfrag{py}[c][c][1.5][0]{$\nu_{q}$ MHz}
\psfrag{px}[c][c][1.5][0]{$\nu_{p}$ MHz}
\psfrag{pz}[c][c][1.5][0]{${\rm Jy}^2\quad\quad$}
\includegraphics[scale=0.74, angle=-90]{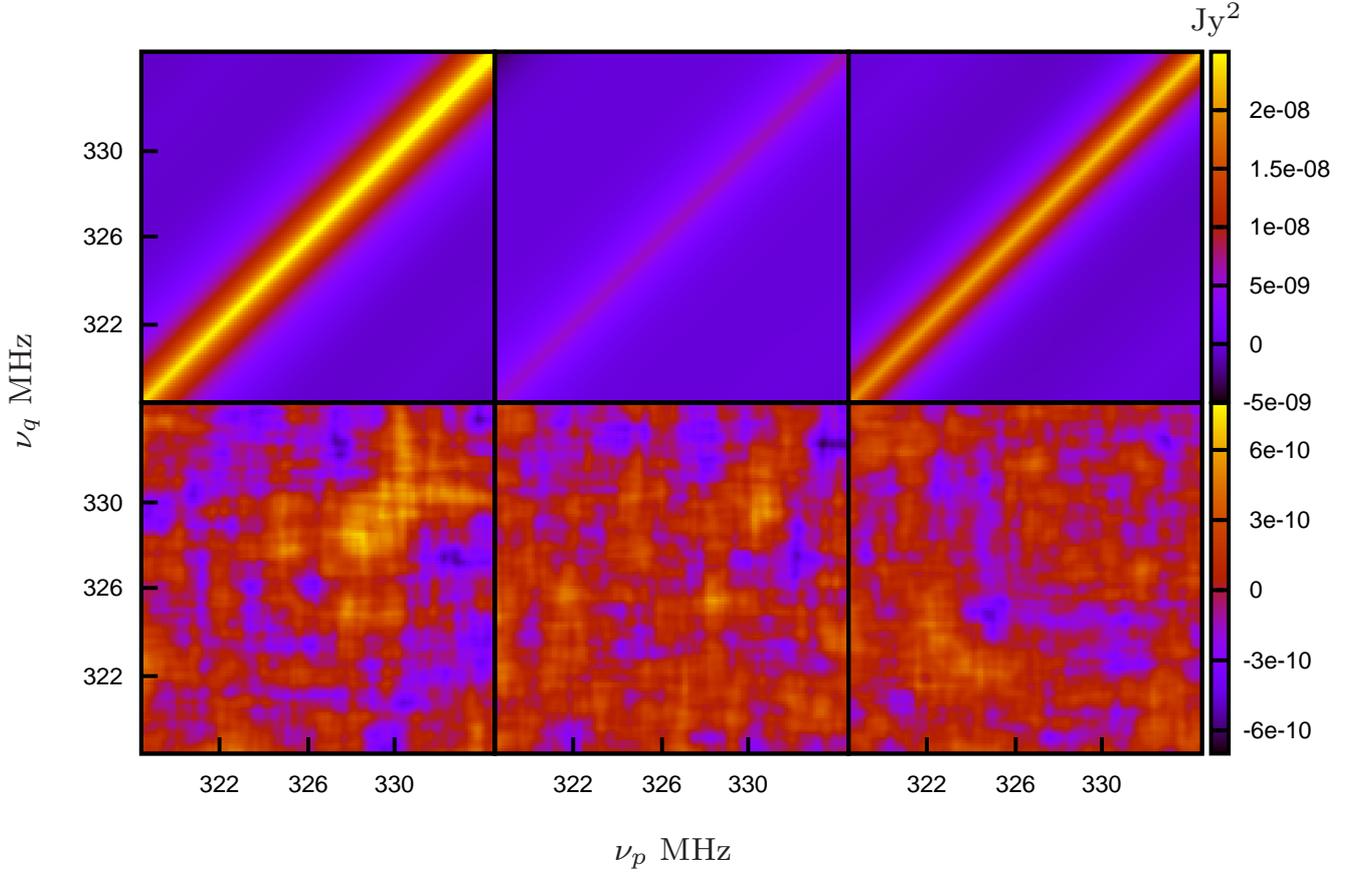}
\caption{The visibility correlation $\mathbfss{S}_2(\mathbfit{U}_m,\nu_p;\mathbfit{U}_n,\nu_q)$ as a
  function of the frequencies $\nu_p$ and $\nu_q$ for the baseline pairs
  $(U_m,U_n)=(U_1,U_1)$ (left), $(U_1,U_2)$ (middle)
  and $(U_2,U_2)$ (right). The upper row shows the theoretical
  predictions which were used as inputs for the simulation and the
  lower row shows its difference with the results averaged over $10^4$
  realizations (color online).}
\label{fig:sigcov}
\end{center}
\end{figure*}

The first step in our simulation is to calculate the expected
visibility correlation matrix $\mathbfss{S}_{2ab}$ (eqs.~\ref{eq:9} and \ref{eq:a3}) of dimensions $4992 \times 4992$.
As discussed earlier, it is convenient to express the OWFA
visibility correlation\footnote{The simulation code is available at https://github.com/anjan-sarkar/eigen-21cm} as $\mathbfss{S}_2(\mathbfit{U}_m,\nu_p;\mathbfit{U}_n,\nu_q)$
(\ref{eq:c2}) for which the non-zero elements can be decomposed
into blocks.  We have a non-zero correlation only when $\mathbfit{U}_m=\mathbfit{U}_n$ (same
baseline) or $\mathbfit{U}_m=\mathbfit{U}_{n \pm 1}$ (adjacent baselines), all the other
correlations are zero. The upper row of Fig. \ref{fig:sigcov} shows
the predicted visibility correlations, the left panel shows
$\mathbfss{S}_2(\mathbfit{U}_1,\nu_p;\mathbfit{U}_1,\nu_q)$ which is the correlation between the
visibilities measured at different frequencies for the fixed baseline
$\mathbfit{U}_1$, the right panel shows the same for $\mathbfit{U}_2$ whereas the middle
panel shows the cross-correlation between the visibilities measured at
$\mathbfit{U}_1$ and  $\mathbfit{U}_2$. Each panel here is a $128 \times 128$
block and there are a total of
$39$ such blocks where the two baselines are the same $(U_m=U_n)$ and
$38$ blocks for the adjacent baseline pairs $(\mathbfit{U}_m=\mathbfit{U}_{n \pm
  1})$. The main features highlighted in the
subsequent discussion are common to all the non-zero blocks including
those not shown here.

 We see that for every baseline pair the visibility correlation peaks
 along the diagonal ($\nu_p = \nu_q)$. The visibility correlation is
 also seen to increase with frequency $\nu_p$ along the diagonal. The
 visibility correlation goes down as we move away from the diagonal
 ($|\nu_p - \nu_q| > 0$), and the signal decorrelates within a frequency
 separation of $ \sim 4 \, {\rm MHz}$. As expected, the visibility
 correlation for the adjacent baseline pair $(\mathbfit{U}_1,\mathbfit{U}_2)$ is
 considerably lower compared to that for the same baselines
 $(\mathbfit{U}_1,\mathbfit{U}_1)$ and $(\mathbfit{U}_2,\mathbfit{U}_2)$. We further observe that the magnitude of
 the visibility correlation decreases as the baseline is increased
 from $(\mathbfit{U}_1,\mathbfit{U}_1)$ to $(\mathbfit{U}_2,\mathbfit{U}_2)$, and it decreases even further for
 the larger baselines.

\begin{figure*}
\begin{center}
\psfrag{py}[c][c][1.2][0]{$\mathbfss{S}_2(\mathbfit{U}_m, \nu_p; \mathbfit{U}_n, \nu_q)$
  Jy$^{2}$} \psfrag{px}[c][c][1.2][0]{$\nu_{p}$ MHz}
\includegraphics[scale=0.435, angle=-90]{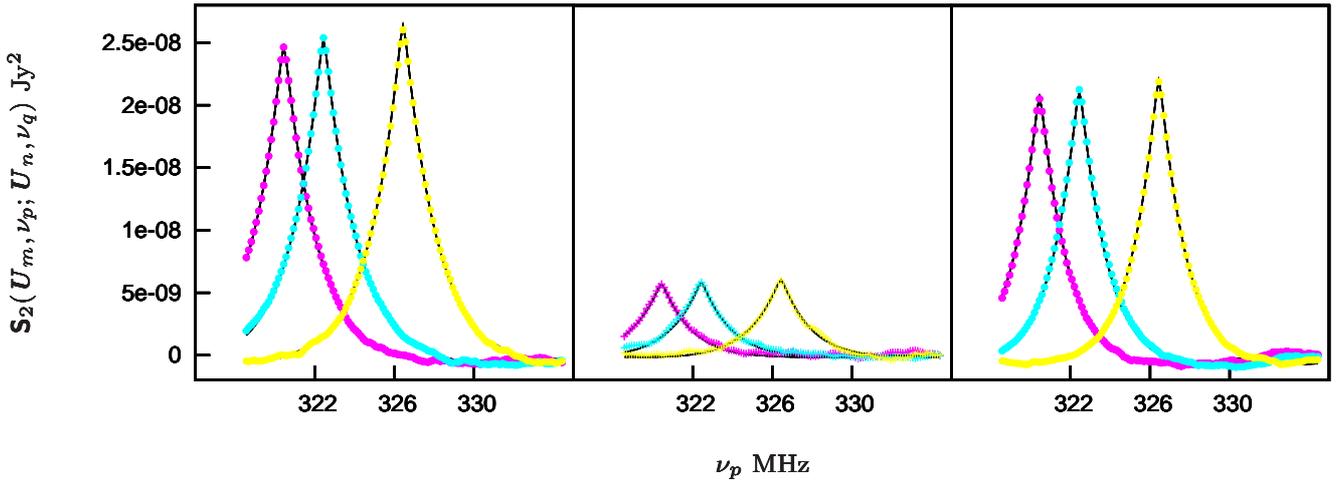}
\caption{$\mathbfss{S}_2(\mathbfit{U}_m, \nu_p; \mathbfit{U}_n,
  \nu_q)$ at a few fixed values of $\nu_q$, shown as solid line plots here, are rows
  from the matrices of the upper panel of Fig. \ref{fig:sigcov}. The points are
  obtained from averaging $10^4$ simulations. The correlation peaks
  when $\nu_p=\nu_q$ (color online).}
\label{fig:sigcovnu}
\end{center}
\end{figure*}

We shall now look at how the visibility correlation
$\mathbfss{S}_2(\mathbfit{U}_m, \nu_p; \mathbfit{U}_n, \nu_q)$ varies with frequency in more
detail. We first consider the variation with frequency separation
$\Delta \nu=\nu_p - \nu_q$.  The curves in the three panels of Fig.
\ref{fig:sigcovnu} show the values of $\mathbfss{S}_2(\mathbfit{U}_m, \nu_p; \mathbfit{U}_n,
\nu_q)$ along different horizontal sections through the corresponding
panels of Fig. \ref{fig:sigcovnu}.  Each curve corresponds to a
fixed value of $\nu_q$ and it shows how the correlation changes as
$\nu_p$ is varied.  We see that the correlation peaks when
$\nu_p=\nu_q$ and falls sharply on either side as $\mid \nu_p - \nu_q
\mid$ increases.  For the baseline pair $(\mathbfit{U}_1,\mathbfit{U}_1)$ the visibility
correlation drops to $0.5$ times the peak value at a frequency
separation of $\Delta \nu_{1/2} \sim 1.1 \, {\rm MHz}$, and drops to
zero at a frequency separation of $\sim 4 \, {\rm MHz}$. The
correlation shows a small amplitude oscillation around zero at larger
frequency separations. The variation with $\nu_p - \nu_q$ is very
similar for the other baseline pairs shown in Fig.
\ref{fig:sigcovnu}.  It is however important to note that the $\Delta
\nu$ dependence changes at the larger baselines $(U > 30)$ where
$\mathbfss{S}_2(\mathbfit{U}_m, \nu_p; \mathbfit{U}_n, \nu_q)$ decorrelates considerably
faster as $\Delta \nu$ is increased (Fig. 7. of
\citealt{ali-bharadwaj2014}).  As noted earlier, the amplitude of the
correlation drops as the baseline pair gets longer from $(\mathbfit{U}_1,\mathbfit{U}_1)$ to
$(\mathbfit{U}_2,\mathbfit{U}_2)$ and it drops by a factor of $\sim 4$
for $(\mathbfit{U}_1,\mathbfit{U}_2)$.

\begin{figure*}
\begin{center}
\vskip.2cm \psfrag{px2}[c][c][1.5][0]{$\mathbfss{S}_2(\mathbfit{U}_m, \nu_p; \mathbfit{U}_n,
  \nu_p)$ Jy$^{2}$} \psfrag{px}[c][c][1.2][0]{$\nu_p$ MHz}
\psfrag{px3}[c][c][1.2][0]{$\mathbfit{U}_m$} \psfrag{py4}[c][c][1.5][0]{}
\psfrag{p2}[c][c][1.0][0]{adjacent} \psfrag{p1}[c][c][1.0][0]{self}
\centerline{{\includegraphics[scale =.7]{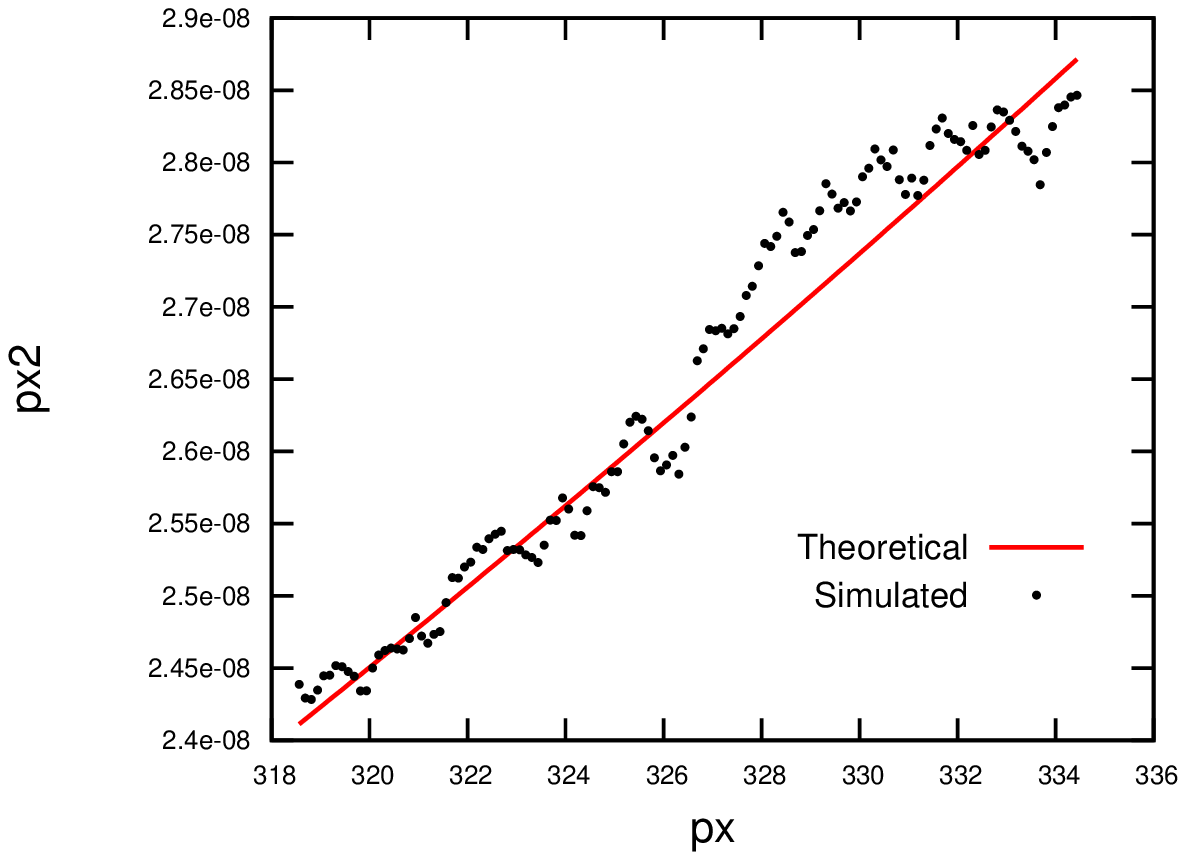}} {
    \includegraphics[scale =.73]{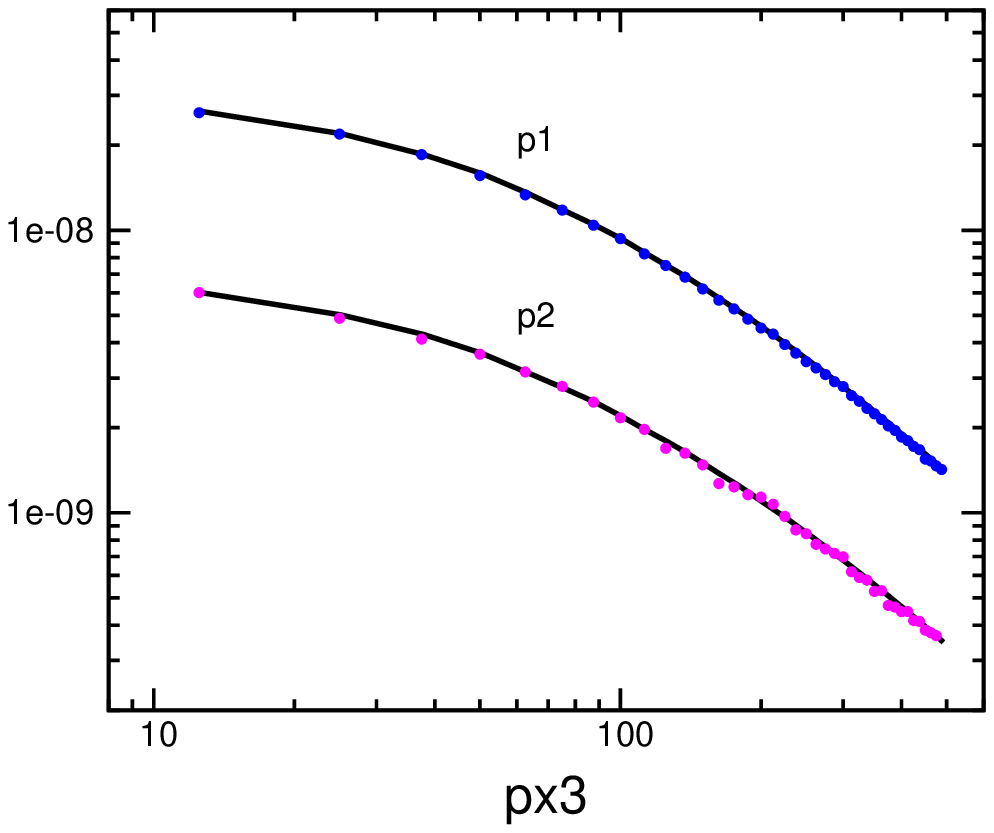}}}
\caption{The left panel shows the variation of $\mathbfss{S}_2$ with $\nu_p$ for the fixed baseline
  pair $(\mathbfit{U}_1,\mathbfit{U}_1)$.  The right panel shows its variation with $\mathbfit{U}_m$
  for $n=m$ (self) and $n=m+1$ (adjacent) with the frequency fixed
  at a value $\nu_p=326.5 \, {\rm MHz}$. In both panels the solid line
  shows the prediction and the points show the mean from
  $10^4$ realizations (color online).}
\label{fig:sigcovbn}
\end{center}
\end{figure*}

The correlations $\mathbfss{S}_2(\mathbfit{U}_m, \nu_q; \mathbfit{U}_n, \nu_p)$ shown in
Fig. \ref{fig:sigcovnu} all peak when $\nu_q=\nu_p$, we now discuss
how the peak value varies with the frequency $\nu_p$. Note that in all
the panels of Fig. \ref{fig:sigcovnu}, the peak value of the
correlation increases as $\nu_p$ is increased.  This is further
illustrated in the left panel of Fig. \ref{fig:sigcovbn} which shows
how the peak correlation changes with $\nu_p$ for the baseline pair
$(\mathbfit{U}_1,\mathbfit{U}_1)$. We see that the predicted
$\mathbfss{S}_2(\mathbfit{U}_1, \nu_p; \mathbfit{U}_1, \nu_p)$ increases linearly by $\sim
20 \%$ as $\nu_p$ is varied across the $B=16 \, {\rm MHz}$ OWFA
bandwidth. This variation of $\mathbfss{S}_2(\mathbfit{U}_1, \nu_p; \mathbfit{U}_1, \nu_p)$
is caused (\ref{eq:a3}) by a combination of (i) the redshift
dependence of the \ion{H}{I} model as quantified by the parameters $Q_{\nu}$
and $\beta$ (ii) the redshift dependence of the comoving distance
$r_{\nu}$, and (iii) the chromatic response of the telescope as
introduced through the variation of the beam pattern and the baseline
with frequency.  We note that $Q_{\nu}$ and $\beta$ change by $\sim
2.5 \%$ and $\sim 0.5 \%$ respectively whereas $r_{\nu}$ varies by
$\sim 2 \%$ over the OWFA bandwidth. This indicates that the frequency
dependence seen in the left panel of Fig. \ref{fig:sigcovbn} is
largely a consequence of the telescope's chromatic response.

The right panel of Fig. \ref{fig:sigcovbn} shows how the visibility
correlation $\mathbf{S}_2(\mathbfit{U}_m, \nu_p; \mathbfit{U}_n, \nu_p)$ varies with
baseline $\mathbfit{U}_m$ for $n=m$ (self) and $n=m + 1$ (adjacent) with the
frequency fixed at a value $\nu_p=326.5 \, {\rm MHz}$.  The other
frequencies not shown here exhibit a similar behavior.  We find that
both the self and adjacent correlation decrease with increasing
baseline $\mathbfit{U}_m$.  This decrease is relatively slow for small baselines
$U < 100$ beyond which it falls faster as $|\mathbfit{U}|^{-1.2}$.  The slope of
the visibility correlation is related to that of the power spectrum
$P(k)$.  We note the baseline $U=100$ corresponds to the Fourier mode
$\mathbfit{k}_{\perp}=2 \pi \mathbfit{U}/r \sim 0.1 \, {\rm Mpc}^{-1}$ where the power
spectrum has a slope $P(k)\sim k^{-1.94}$.  We see that the self and
adjacent correlations show a very similar $\mathbfit{U}_m$ dependence, the
adjacent correlations are a factor of $\sim 4.3$ smaller than the self
correlation at the smallest baselines and this factor changes to
$\sim 4$ at the largest baselines.

\begin{figure}
\begin{center}
\psfrag{py}[c][c][1.2][0]{$\mathbfit{S}(\mathbfit{U}_n, \nu) \times 10^{-5}$ Jy}
\psfrag{px}[c][c][1.5][0]{$\nu$ MHz} \psfrag{p1}[c][c][1.2][0]{$n =
  1$} \psfrag{p16}[c][c][1.2][0]{$n = 16$}
\psfrag{p32}[c][c][1.2][0]{$n = 32$} \includegraphics[scale=1.05,
  angle=-90]{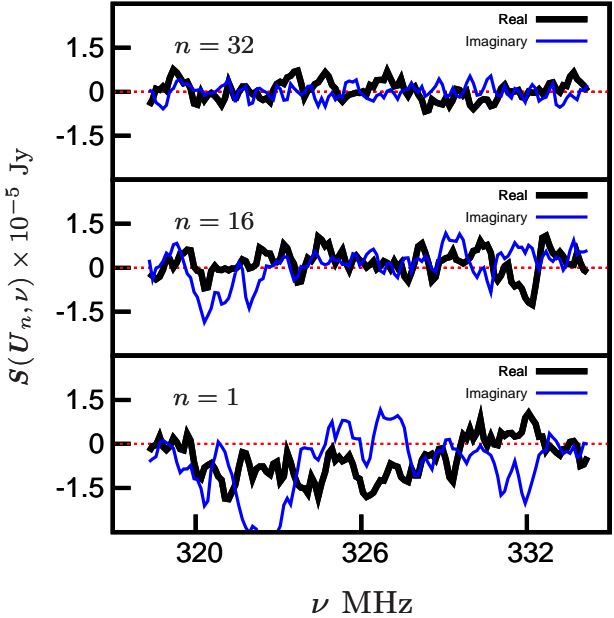}
\caption{The real and imaginary parts of a single
  realization of the simulated visibilities for the fixed baselines
  $\mathbfit{U}_n$ with $n=1,\, 16 \, {\rm and} \, 32$ respectively. The dotted
  line in each panel is the expectation of the visibilities, $\langle
  \mathbfit{S}(\mathbfit{U}_n, \nu) \rangle = 0$ (color online).}
\label{fig:sigvis}
\end{center}
\end{figure}

In the next step, we have determined the eigenvalues and eigenvectors
of the predicted visibility correlation $\mathbfss{S}_{2ab}$ to
simulate (\ref{eq:a7}) different realizations of the
visibilities. For three different baselines, Fig. \ref{fig:sigvis}
shows the real and imaginary parts for a single realization of the
simulated visibilities. As expected, the visibilities are random
fluctuating quantities with zero mean.  We see that the visibilities
at the smallest baseline $\mathbfit{U}_1$ fluctuate relatively slowly with
varying frequency as compared to the larger baselines where the signal
fluctuates more rapidly with varying frequency.  This is consistent
with the theoretical input that the visibilities at the smaller
baselines remain correlated over a larger frequency separation as
compared to the visibilities at the larger baselines.

A total of $10^4$ Gaussian random realizations of the
visibilities $\mathbfit{S}_a$ were simulated. As mentioned earlier, a single random
  realization of the visibilities can be simulated (eqs.~\ref{eq:a7}
  and \ref{eq:a5}) using a set of $2N$ Gaussian random variables
  $(x_{\alpha},y_{\alpha})$. We then averaged over these
simulations to estimate the mean visibility correlation for OWFA. The lower row of Fig. \ref{fig:sigcov} shows the difference
  between the simulated and the predicted visibility
  correlations for the same baseline pairs as those for which the
  theoretical predictions have been shown in the upper row. We find
  that difference between the simulated and the theoretical
  predictions are quite small, which indicates that our
  simulation faithfully reproduces the theoretically
  predicted visibility covariance. The points in Fig.
\ref{fig:sigcovnu} and Fig. \ref{fig:sigcovbn} show the visibility
correlation values obtained from the simulations: the results from simulations,
shown as the points, are found to be in excellent agreement with the predictions
shown as the solid curves. The deviations between the
  simulation and the theory seen in the left panel of
  Fig. \ref{fig:sigcovbn} can be attributed to the statistical
  fluctuations inherent to the signal. The coherence scale in frequency in
  Fig. \ref{fig:sigcovbn} reflects the frequency correlation seen
  in Fig. \ref{fig:sigcovnu}. 

\begin{figure*}
\begin{center}
\psfrag{py}[c][c][1.5][0]{$\nu_{q}$ MHz}
\psfrag{px}[c][c][1.5][0]{$\nu_{p}$ MHz}
\psfrag{pz}[c][c][1.5][-90]{$r$} \includegraphics[scale=0.75,
  angle=-90]{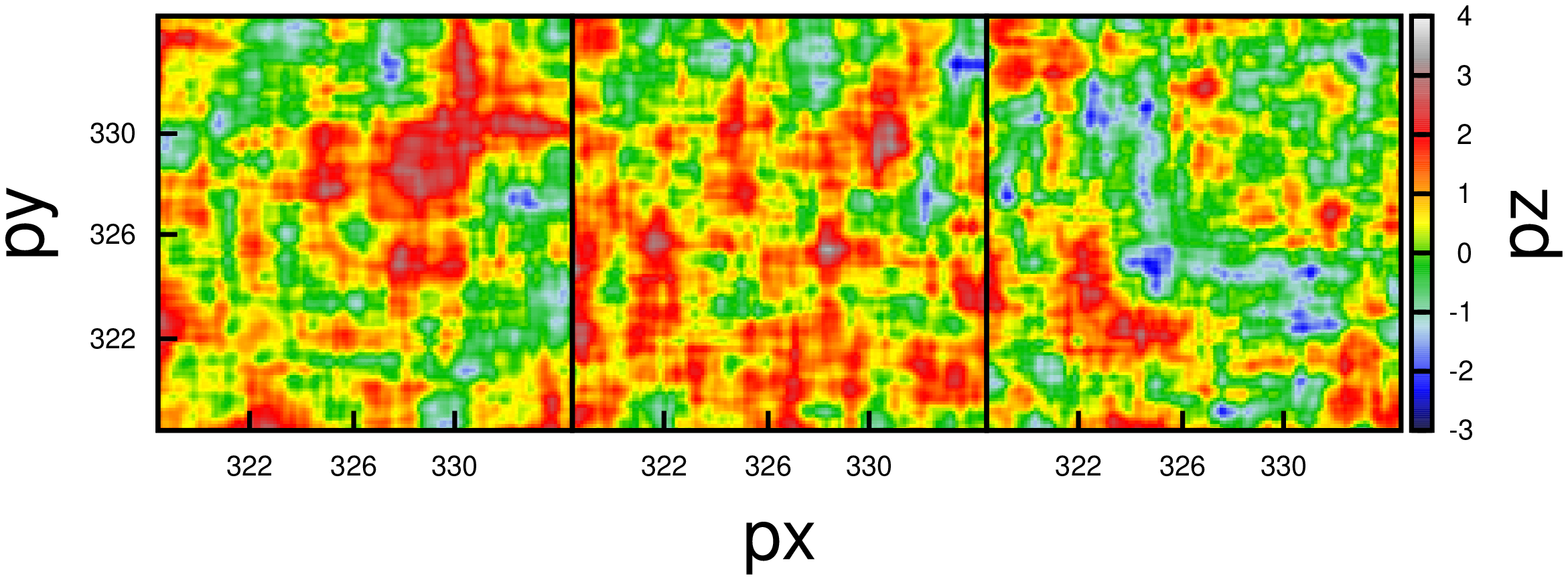}
\caption{Each panel shows $r$ (defined in \ref{eq:d1}), which is a dimensionless
  ratio that quantifies the difference between the simulated and the
  predicted visibility correlations, for each baseline pair of
  Fig. \ref{fig:sigcov} (color online). 
}
\label{fig:sigcovvarcm}
\end{center}
\end{figure*}

\begin{figure*}
\centering
\begin{minipage}{190mm}
  \psfrag{py}[c][c][1.2][0]{$\Delta N/N$} \psfrag{px}[c][c][1.5][0]{$r$}
  \psfrag{p1}[c][c][1.2][0]{$n = 1$} \psfrag{p16}[c][c][1.2][0]{$n =
   16$} \psfrag{p32}[c][c][1.2][0]{$n = 32$}
\subfigure[$\mathbfit{U}_1-\mathbfit{U}_1$]{\includegraphics[scale=0.44]{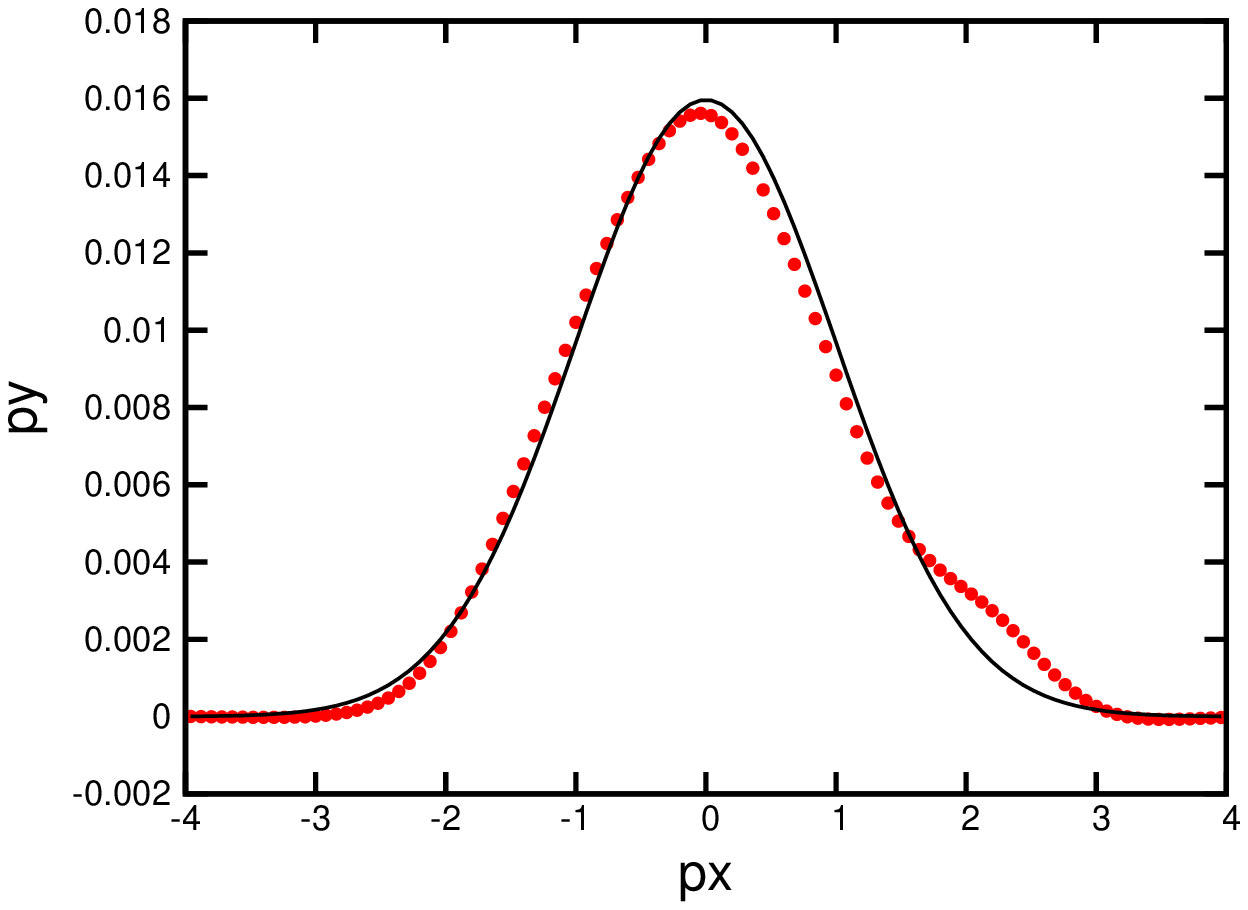}}
\hskip 4mm
\subfigure[$\mathbfit{U}_1-\mathbfit{U}_2$]{\includegraphics[scale=0.44]{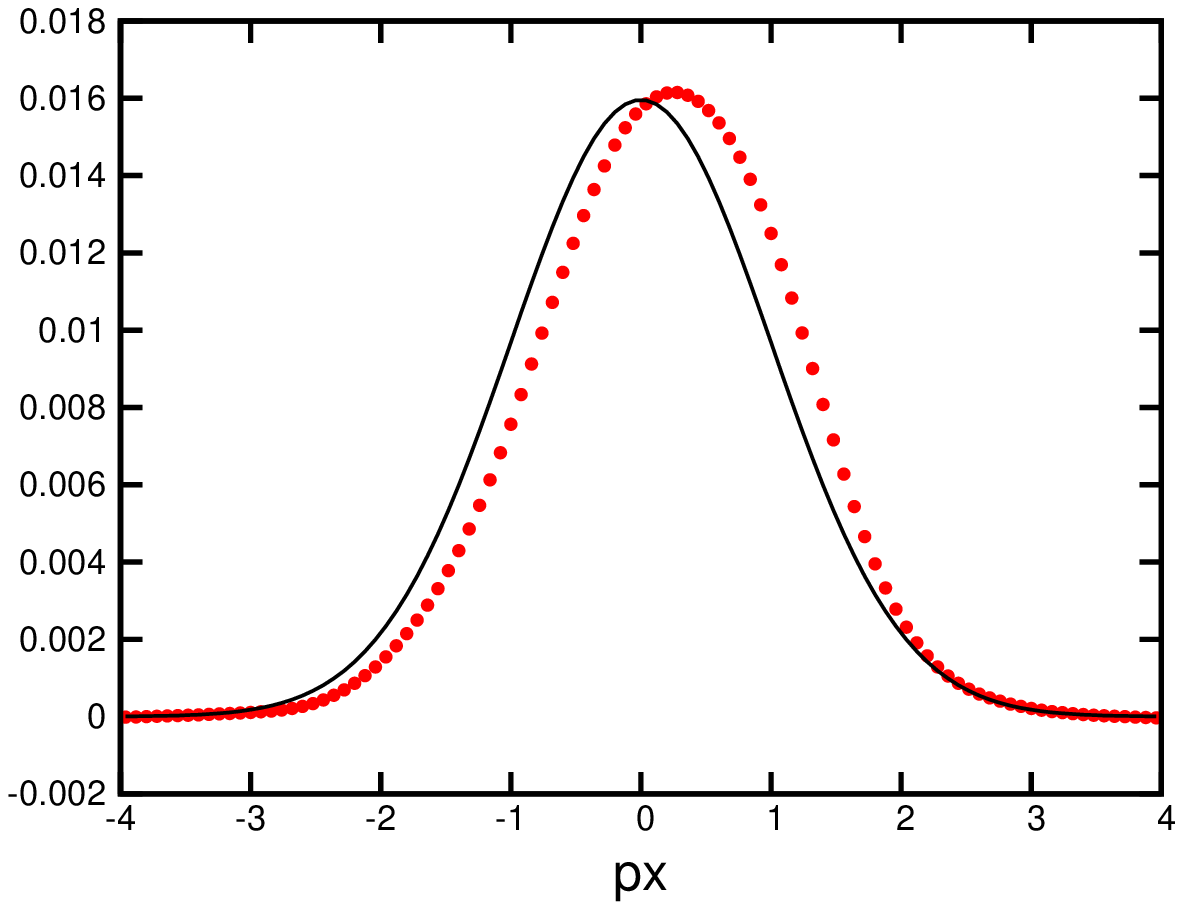}}
\hskip 4mm
\subfigure[$\mathbfit{U}_2-\mathbfit{U}_2$]{\includegraphics[scale=0.44]{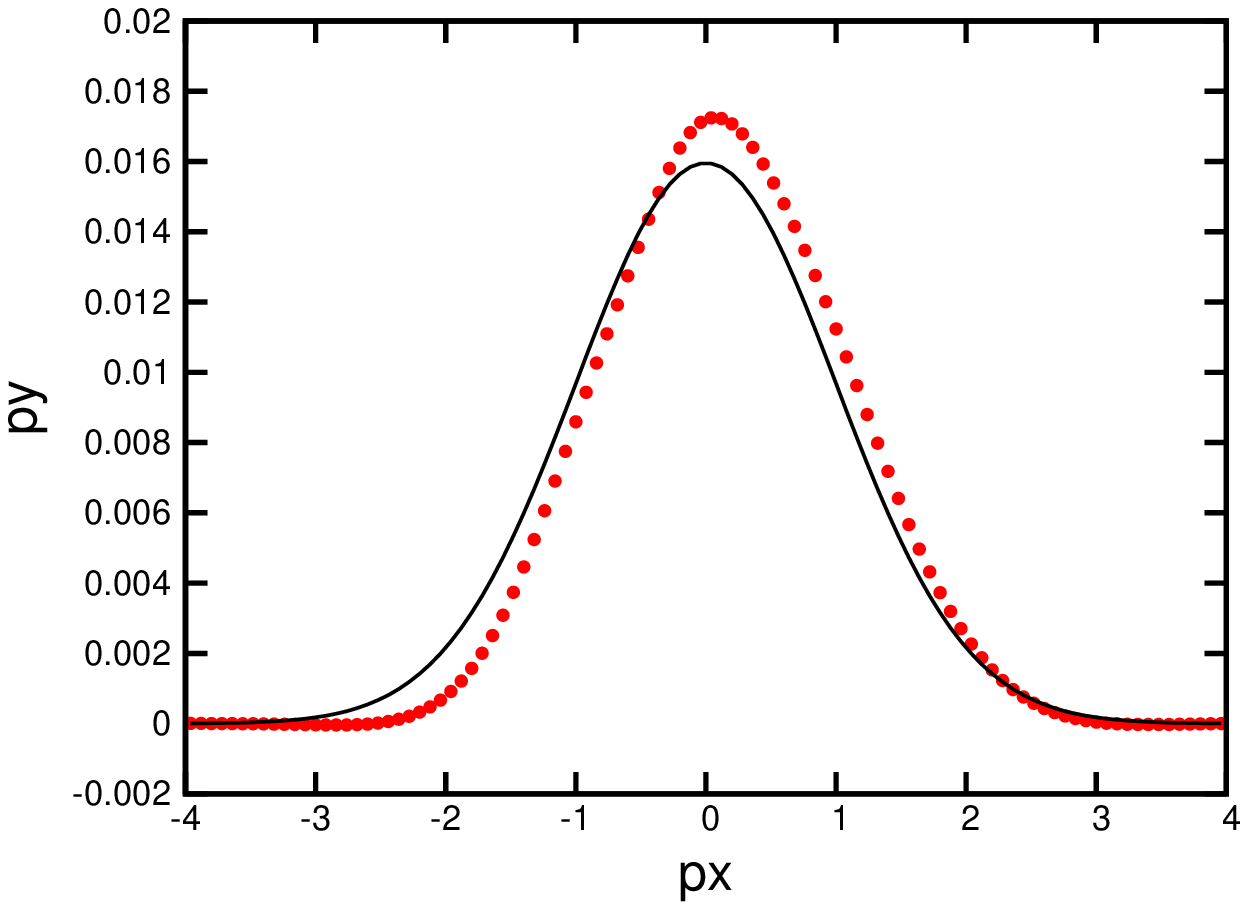}}
\end{minipage}
\caption{The points show the distribution of the values of $r$ from
  Fig. \ref{fig:sigcovvarcm}.  The $r$ values have
  been binned with an interval $\Delta r=0.04$ and the solid line shows
  the prediction for a Gaussian of unit variance. Only every other point in the
  binned histogram has been shown for clarity (color online). }
\label{fig:sigfn}
\end{figure*}

We now make a quantitative comparison between the
simulated and the predicted visibility correlations, for which we use the ratio
\begin{equation}
r=\frac{\Delta \mathbfss{S}_2 \times \sqrt{N_r}}{\sigma_{\mathbfss{S}_2}}
\label{eq:d1}
\end{equation}
where $\Delta \mathbfss{S}_2$ is the difference between the simulated and the
predicted visibility correlation
$\mathbfss{S}_2(\mathbfit{U}_m,\nu_p;\mathbfit{U}_n,\nu_q)$, $N_r$ is the number of realizations of
the simulations and $\sigma_{\mathbfss{S}_2}^2$ is the
predicted variance for the visibility correlation.
Under the assumption that the visibilities are Gaussian random
variables, we have calculated the variance to be
\begin{flalign}
  \sigma^2_{\mathbfss{s}_2} = \frac{1}{2} [\mathbfss{S}^2_2(\mathbfit{U}_m, \nu_p; \mathbfit{U}_n,
    \nu_q) \nonumber  + \mathbfss{S}_2(\mathbfit{U}_m, \nu_p; \mathbfit{U}_m,
    \nu_p) & \\
    \times\ \mathbfss{S}_2(\mathbfit{U}_n, \nu_q; \mathbfit{U}_n,
    \nu_q)].
\label{eq:d2}
\end{flalign}
The dimensionless ratio $r$ quantifies the difference between the
simulated and predicted visibility correlations. We
expect the values of $r$ to have a Gaussian distribution of unit
variance and zero mean. Fig. \ref{fig:sigcovvarcm} shows $r$ for the
baselines pairs of Fig. \ref{fig:sigcov}, while 
Fig. \ref{fig:sigfn} shows the statistics  of the values of $r$.
We see that this is in  
reasonably good agreement with a Gaussian distribution of unit
variance. The deviations seen can be attributed to the inherent
random fluctuations of the signal. The deviations between the
  predicted and the simulated distribution are of the order of a
few  percent, which is expected with $10^4$ Gaussian random
  realizations. 

\section{Discussion and summary}
The 21-cm intensity signal is buried in foregrounds which
  are several orders of magnitude larger. Precision of calibration, the
  telescope's chromatic response, beam and other instrumental
  systematics pose additional challenges.  Simulations play a vital role
in testing and validating \ion{H}{I} 21-cm power spectrum estimation
techniques. Simulations are particularly important in quantifying the
impact of foreground removal, calibration, and various instrumental
and post-processing effects.  Conventional methods for simulating the
expected 21-cm visibility signal requires simulating the sky signal
which is then passed through a software model of the instrument to
generate the visibilities.  This may pose a computational challenge in
that it is necessary to simulate the \ion{H}{I} distribution in a large
computational volume. Further, the \ion{H}{I} signal and the cosmological
parameters both vary with line of sight distance within the
observational volume due to the light cone effect.  The computational
requirements scale in such cases with the number of independent realizations of
the signal sought. In this paper, we present an analytical
method to simulate the \ion{H}{I} visibility signal
and have demonstrated applying it to simulate the \ion{H}{I} visibility signal for the upcoming
OWFA Phase I.

The first step in our method is to compute the expected visibility
correlation for the signal at different baselines and frequency
channels. Our approach for calculating the visibility correlation
differs from the previous studies on two accounts. First, we have
incorporated the light-cone effect whereby the statistical properties
of the \ion{H}{I} signal and cosmological parameters both evolve with redshift
or equivalently, with observing frequency. Second, we have also
accounted for the fact the telescope has a chromatic response, 
  that is the telescope parameters and baselines vary
with observing frequency. The combined effect of these two features
makes the \ion{H}{I} visibility signal non-ergodic in frequency, which
  implies that the correlation between the visibilities at two
different frequencies $\nu_p$ and $\nu_q$ is no longer a function of
the frequency separation $\nu_p - \nu_q$. The Fourier basis or the
delay channel $\tau$ cease to be the appropriate choice once this
ergodicity is broken.  We note that the general formalism adopted here
does not assume the \ion{H}{I} visibility signal to be ergodic in
frequency. We find that the deviations from ergodicity are $\sim 20
\%$ for the limited bandwidth of $16 \, {\rm MHz}$ considered here,
however these effects are expected to be important for observations
that span a larger bandwidth.

As was shown by Kosambi \citep{Kosambi1943} and subsequently Karhunen \citep{karhunen1947} and Lo{\'e}ve \citep{loeve1955}, the
eigenvectors of the visibility correlation matrix provide the correct
basis for analyzing the \ion{H}{I} visibility signal even if the ergodicity is
broken. We have used the eigenvalues and eigenvectors of the
predicted visibility correlation matrix to simulate the
\ion{H}{I} visibility signal.  The computational effort here goes into
calculating the expected visibility correlation matrix, and finding
its eigenvalues and eigenvectors. Once the eigenvalues and
eigenvectors are known, it is only necessary to generate a set of
random numbers (\ref{eq:a7}) to simulate an entire set of the \ion{H}{I}
visibility signal. Multiple random
realizations of the \ion{H}{I} visibility signal can be simulated by using different
realizations of the random numbers. This feature makes this method
particularly efficient if one wishes to simulate many random
realizations of the \ion{H}{I} visibility signal. In this paper we have
simulated $10^4$ random realizations of the \ion{H}{I} visibility signal and
used these to calculate the visibility correlation matrix. We find
that the simulations are in very good agreement with the theoretical
predictions which have been used as input.

The simulation method presented here is particularly well suited for
OWFA which has only a few independent baselines that do not change
with the rotation of the Earth. The situation could become significantly
more complex for an array like the GMRT which has many different
baselines that change with Earth rotation. In such cases, conventional
techniques which involve simulating the sky signal may be
more efficient. However, we note that 
this limitation could be overcome by gridding the baselines to
reduce the complexity of the problem.  Another limitation of the
present technique is that it assumes the 21-cm signal to be a Gaussian
random field ignoring the non-Gaussianity which would arise due to the
non-linear evolution of the underlying matter perturbations. This
aspect of the 21-cm signal would be naturally incorporated in a N-body
simulation. It is useful to
note that the method presented here can be extended to handle a
\ion{H}{I} signal field that is not purely Gaussian: non-Gaussianity can be incorporated by modifying the statistics of
the random variables used to simulate the \ion{H}{I} visibility signal, which we
plan to address in the future.

\section{acknowledgment}
AKS thanks Debanjan Sarkar, Suman Chatterjee and Rajesh
Mondal for their help in preparing the figures. AKS makes a special mention of
Debanjan Sarkar who read through the draft of the
manuscript in detail and suggested useful changes. VRM thanks CTS-IIT KGP for
hosting a visit which led to the genesis of this work. The authors are 
grateful to the reviewers of the manuscript for their critical views and
suggestions, which resulted in a substantial improvement to the clarity of the
presentation.


\bibliographystyle{mnras}
\bibliography{ref} 

\begin{thebibliography}{}
\makeatletter
\relax
\def\mn@urlcharsother{\let\do\@makeother \do\$\do\&\do\#\do\^\do\_\do\%\do\~}
\def\mn@doi{\begingroup\mn@urlcharsother \@ifnextchar [ {\mn@doi@}
  {\mn@doi@[]}}
\def\mn@doi@[#1]#2{\def\@tempa{#1}\ifx\@tempa\@empty \href
  {http://dx.doi.org/#2} {doi:#2}\else \href {http://dx.doi.org/#2} {#1}\fi
  \endgroup}
\def\mn@eprint#1#2{\mn@eprint@#1:#2::\@nil}
\def\mn@eprint@arXiv#1{\href {http://arxiv.org/abs/#1} {{\tt arXiv:#1}}}
\def\mn@eprint@dblp#1{\href {http://dblp.uni-trier.de/rec/bibtex/#1.xml}
  {dblp:#1}}
\def\mn@eprint@#1:#2:#3:#4\@nil{\def\@tempa {#1}\def\@tempb {#2}\def\@tempc
  {#3}\ifx \@tempc \@empty \let \@tempc \@tempb \let \@tempb \@tempa \fi \ifx
  \@tempb \@empty \def\@tempb {arXiv}\fi \@ifundefined
  {mn@eprint@\@tempb}{\@tempb:\@tempc}{\expandafter \expandafter \csname
  mn@eprint@\@tempb\endcsname \expandafter{\@tempc}}}

\bibitem[\protect\citeauthoryear{Ade et~al.,}{Ade et~al.}{2014}]{ade2014}
Ade P.,  et~al., 2014, \aap, 571, A16

\bibitem[\protect\citeauthoryear{Aguirre et~al.,}{Aguirre
  et~al.}{2014}]{aguirre2014}
Aguirre J.~E.,  et~al., 2014, in Radio Science Meeting (USNC-URSI NRSM), 2014
  United States National Committee of URSI National. pp~1--1

\bibitem[\protect\citeauthoryear{Ali \& Bharadwaj}{Ali \&
  Bharadwaj}{2014}]{ali-bharadwaj2014}
Ali S.~S.,  Bharadwaj S.,  2014, \japa, 35, 157

\bibitem[\protect\citeauthoryear{Ali, Bharadwaj  \& Pandey}{Ali
  et~al.}{2005}]{ali-pandey2005}
Ali S.~S.,  Bharadwaj S.,   Pandey B.,  2005, \mnras, 363, 251

\bibitem[\protect\citeauthoryear{Ansari, Campagne, Colom, Magneville, Martin,
  Moniez, Rich  \& Yeche}{Ansari et~al.}{2011}]{ansari2011}
Ansari R.,  Campagne J.,  Colom P.,  Magneville C.,  Martin J.,  Moniez M.,
  Rich J.,   Yeche C.,  2011, arXiv preprint arXiv:1106.5659

\bibitem[\protect\citeauthoryear{Ansari et~al.,}{Ansari
  et~al.}{2012}]{ansari2012}
Ansari R.,  et~al., 2012, \aap, 540, A129

\bibitem[\protect\citeauthoryear{Bandura et~al.,}{Bandura
  et~al.}{2014}]{bandura2014}
Bandura K.,  et~al., 2014, in SPIE Astronomical Telescopes+ Instrumentation. pp
  914522--914522

\bibitem[\protect\citeauthoryear{Battye et~al.,}{Battye
  et~al.}{2012}]{battye2012}
Battye R.,  et~al., 2012, arXiv:1209.1041

\bibitem[\protect\citeauthoryear{Battye et~al.,}{Battye
  et~al.}{2016}]{battye2016}
Battye R.,  et~al., 2016, arXiv:1610.06826

\bibitem[\protect\citeauthoryear{Bharadwaj \& Ali}{Bharadwaj \&
  Ali}{2005}]{bharadwaj-ali2005}
Bharadwaj S.,  Ali S.~S.,  2005, \mnras, 356, 1519

\bibitem[\protect\citeauthoryear{Bharadwaj \& Pandey}{Bharadwaj \&
  Pandey}{2003}]{bharadwaj2003}
Bharadwaj S.,  Pandey S.~K.,  2003, \japa, 24, 23

\bibitem[\protect\citeauthoryear{Bharadwaj \& Sethi}{Bharadwaj \&
  Sethi}{2001}]{bharadwaj-sethi2001}
Bharadwaj S.,  Sethi S.~K.,  2001, \japa, 22, 293

\bibitem[\protect\citeauthoryear{Bharadwaj \& Srikant}{Bharadwaj \&
  Srikant}{2004}]{bharadwaj-srikant2004}
Bharadwaj S.,  Srikant P.~S.,  2004, \japa, 25, 67

\bibitem[\protect\citeauthoryear{Bharadwaj, Nath  \& Sethi}{Bharadwaj
  et~al.}{2001}]{bharadwaj2001}
Bharadwaj S.,  Nath B.~B.,   Sethi S.~K.,  2001, \japa, 22, 21

\bibitem[\protect\citeauthoryear{Bharadwaj, Sethi  \& Saini}{Bharadwaj
  et~al.}{2009}]{bharadwaj2009}
Bharadwaj S.,  Sethi S.~K.,   Saini T.~D.,  2009, \prd, 79, 083538

\bibitem[\protect\citeauthoryear{Bharadwaj, Sarkar  \& Ali}{Bharadwaj
  et~al.}{2015}]{bharadwaj-fisher2015}
Bharadwaj S.,  Sarkar A.~K.,   Ali S.~S.,  2015, \japa, 36, 385

\bibitem[\protect\citeauthoryear{Bowman, Morales  \& Hewitt}{Bowman
  et~al.}{2009}]{bowman2009}
Bowman J.~D.,  Morales M.~F.,   Hewitt J.~N.,  2009, \apj, 695, 183

\bibitem[\protect\citeauthoryear{Bull, Camera, Raccanelli, Blake, Ferreira,
  Santos  \& Schwarz}{Bull et~al.}{2015a}]{bull2015}
Bull P.,  Camera S.,  Raccanelli A.,  Blake C.,  Ferreira P.~G.,  Santos M.~G.,
    Schwarz D.~J.,  2015a, arXiv:1501.04088

\bibitem[\protect\citeauthoryear{Bull, Ferreira, Patel  \& Santos}{Bull
  et~al.}{2015b}]{bull-ferreira2015}
Bull P.,  Ferreira P.~G.,  Patel P.,   Santos M.~G.,  2015b, \apj, 803, 21

\bibitem[\protect\citeauthoryear{Chang, Pen, Peterson  \& McDonald}{Chang
  et~al.}{2008}]{chang2008}
Chang T.-C.,  Pen U.-L.,  Peterson J.~B.,   McDonald P.,  2008, \prl, 100,
  091303

\bibitem[\protect\citeauthoryear{{Chatterjee}, {Bharadwaj}  \&
  {Marthi}}{{Chatterjee} et~al.}{2017}]{chatterjee2017}
{Chatterjee} S.,  {Bharadwaj} S.,   {Marthi} V.~R.,  2017, \mn@doi [\japa]
  {10.1007/s12036-017-9433-1}, \href
  {http://adsabs.harvard.edu/abs/2017JApA...38...15C} {38, 15}

\bibitem[\protect\citeauthoryear{Chen}{Chen}{2012}]{chen2012}
Chen X.,  2012, in International Journal of Modern Physics: Conference Series.
  pp 256--263

\bibitem[\protect\citeauthoryear{Chengalur, Gupta  \& Dwarkanath}{Chengalur
  et~al.}{2007}]{book-GMRT}
Chengalur J.~N.,  Gupta Y.,   Dwarkanath K.,  2007, Low frequency radio
  astronomy 3rd edition

\bibitem[\protect\citeauthoryear{{Cleary} et~al.,}{{Cleary}
  et~al.}{2016}]{comap16}
{Cleary} K.,  et~al., 2016, in American Astronomical Society Meeting Abstracts.
  p. 426.06

\bibitem[\protect\citeauthoryear{Crites et~al.,}{Crites
  et~al.}{2014}]{crites2014}
Crites A.,  et~al., 2014.

\bibitem[\protect\citeauthoryear{DeBoer et~al.,}{DeBoer
  et~al.}{2017}]{deboer2017}
DeBoer D.~R.,  et~al., 2017, \pasp, 129, 045001

\bibitem[\protect\citeauthoryear{Delhaize, Meyer, Staveley-Smith  \&
  Boyle}{Delhaize et~al.}{2013}]{delhaize2013}
Delhaize J.,  Meyer M.,  Staveley-Smith L.,   Boyle B.,  2013, \mnras, p.
  stt810

\bibitem[\protect\citeauthoryear{Eisenstein \& Hu}{Eisenstein \&
  Hu}{1998}]{eisenstein-hu1998}
Eisenstein D.~J.,  Hu W.,  1998, \apj, 496, 605

\bibitem[\protect\citeauthoryear{Gehlot \& Bagla}{Gehlot \&
  Bagla}{2017}]{gehlot-bagla2017}
Gehlot B.~K.,  Bagla J.,  2017, \japa, 38, 13

\bibitem[\protect\citeauthoryear{Ghosh, Bharadwaj, Ali  \& Chengalur}{Ghosh
  et~al.}{2011a}]{Ghosh2011a}
Ghosh A.,  Bharadwaj S.,  Ali S.~S.,   Chengalur J.~N.,  2011a, \mnras, 411,
  2426

\bibitem[\protect\citeauthoryear{Ghosh, Bharadwaj, Ali  \& Chengalur}{Ghosh
  et~al.}{2011b}]{ghosh2011b}
Ghosh A.,  Bharadwaj S.,  Ali S.~S.,   Chengalur J.~N.,  2011b, \mnras, 418,
  2584

\bibitem[\protect\citeauthoryear{Ghosh, Prasad, Bharadwaj, Ali  \&
  Chengalur}{Ghosh et~al.}{2012}]{ghosh2012}
Ghosh A.,  Prasad J.,  Bharadwaj S.,  Ali S.~S.,   Chengalur J.~N.,  2012,
  \mnras, 426, 3295

\bibitem[\protect\citeauthoryear{Gleser, Nusser  \& Benson}{Gleser
  et~al.}{2008}]{gleser2008}
Gleser L.,  Nusser A.,   Benson A.~J.,  2008, \mnras, 391, 383

\bibitem[\protect\citeauthoryear{Hazra \& Sarkar}{Hazra \&
  Sarkar}{2012}]{hazra2012}
Hazra D.~K.,  Sarkar T.~G.,  2012, \prl, 109, 121301

\bibitem[\protect\citeauthoryear{Jeli{\'c} et~al.,}{Jeli{\'c}
  et~al.}{2008}]{jelic2008}
Jeli{\'c} V.,  et~al., 2008, \mnras, 389, 1319

\bibitem[\protect\citeauthoryear{Kaiser}{Kaiser}{1987}]{kaiser1987}
Kaiser N.,  1987, \mnras, 227, 1

\bibitem[\protect\citeauthoryear{Karhunen}{Karhunen}{1947}]{karhunen1947}
Karhunen K.,  1947, Uber lineare Methoden in der Wahrscheinlichkeitsrechnung.
Universitat Helsinki

\bibitem[\protect\citeauthoryear{Kobayashi, Mark  \& Turin}{Kobayashi
  et~al.}{2011}]{turin2011}
Kobayashi H.,  Mark B.~L.,   Turin W.,  2011, Probability, random processes,
  and statistical analysis: applications to communications, signal processing,
  queueing theory and mathematical finance.
Cambridge University Press

\bibitem[\protect\citeauthoryear{Kosambi}{Kosambi}{1943}]{Kosambi1943}
Kosambi D.,  1943, J. Indian Math. Soc, 7, 76

\bibitem[\protect\citeauthoryear{Lah et~al.,}{Lah et~al.}{2007}]{lah2007}
Lah P.,  et~al., 2007, \mnras, 376, 1357

\bibitem[\protect\citeauthoryear{Lanzetta, Wolfe  \& Turnshek}{Lanzetta
  et~al.}{1995}]{lanzetta1995}
Lanzetta K.~M.,  Wolfe A.~M.,   Turnshek D.~A.,  1995, \apj, 440, 435

\bibitem[\protect\citeauthoryear{Loeb \& Wyithe}{Loeb \&
  Wyithe}{2008}]{loeb-wyithe2008}
Loeb A.,  Wyithe J. S.~B.,  2008, \prl, 100, 161301

\bibitem[\protect\citeauthoryear{Lo{\`e}ve}{Lo{\`e}ve}{1955}]{loeve1955}
Lo{\`e}ve M.,  1955, Probability theory: foundations, random sequences.
van Nostrand Princeton, NJ

\bibitem[\protect\citeauthoryear{Mar{\'\i}n, Gnedin, Seo  \&
  Vallinotto}{Mar{\'\i}n et~al.}{2010}]{marin2010}
Mar{\'\i}n F.~A.,  Gnedin N.~Y.,  Seo H.-J.,   Vallinotto A.,  2010, \apj, 718,
  972

\bibitem[\protect\citeauthoryear{{Marthi}}{{Marthi}}{2017}]{marthi2017}
{Marthi} V.~R.,  2017, \mn@doi [\japa] {10.1007/s12036-017-9429-x}, \href
  {http://adsabs.harvard.edu/abs/2017JApA...38...12M} {38, 12}

\bibitem[\protect\citeauthoryear{{Marthi} \& {Chengalur}}{{Marthi} \&
  {Chengalur}}{2014}]{marthi-chengalur2014}
{Marthi} V.~R.,  {Chengalur} J.,  2014, \mn@doi [\mnras]
  {10.1093/mnras/stt1902}, \href
  {http://adsabs.harvard.edu/abs/2014MNRAS.437..524M} {437, 524}

\bibitem[\protect\citeauthoryear{{Marthi}, {Chatterjee}, {Chengalur}  \&
  {Bharadwaj}}{{Marthi} et~al.}{2017}]{marthi-chatterjee2017}
{Marthi} V.~R.,  {Chatterjee} S.,  {Chengalur} J.~N.,   {Bharadwaj} S.,  2017,
  \mn@doi [\mnras] {10.1093/mnras/stx1796}, \href
  {http://adsabs.harvard.edu/abs/2017MNRAS.471.3112M} {471, 3112}

\bibitem[\protect\citeauthoryear{Martin, Papastergis, Giovanelli, Haynes,
  Springob  \& Stierwalt}{Martin et~al.}{2010}]{martin2010}
Martin A.~M.,  Papastergis E.,  Giovanelli R.,  Haynes M.~P.,  Springob C.~M.,
   Stierwalt S.,  2010, \apj, 723, 1359

\bibitem[\protect\citeauthoryear{McQuinn, Zahn, Zaldarriaga, Hernquist  \&
  Furlanetto}{McQuinn et~al.}{2006}]{mcquinn2006}
McQuinn M.,  Zahn O.,  Zaldarriaga M.,  Hernquist L.,   Furlanetto S.~R.,
  2006, \apj, 653, 815

\bibitem[\protect\citeauthoryear{Meiring et~al.,}{Meiring
  et~al.}{2011}]{meiring2011}
Meiring J.~D.,  et~al., 2011, \apj, 732, 35

\bibitem[\protect\citeauthoryear{Morales}{Morales}{2005}]{Morales2005}
Morales M.~F.,  2005, \apj, 619, 678

\bibitem[\protect\citeauthoryear{Newburgh et~al.,}{Newburgh
  et~al.}{2016}]{newburgh2016}
Newburgh L.,  et~al., 2016, arXiv:1607.02059

\bibitem[\protect\citeauthoryear{Noterdaeme et~al.,}{Noterdaeme
  et~al.}{2012}]{noterdaeme2012}
Noterdaeme P.,  et~al., 2012, \aap, 547, L1

\bibitem[\protect\citeauthoryear{Parsons et~al.,}{Parsons
  et~al.}{2010}]{parsons2010}
Parsons A.~R.,  et~al., 2010, \aj, 139, 1468

\bibitem[\protect\citeauthoryear{Parsons, Pober, Aguirre, Carilli, Jacobs  \&
  Moore}{Parsons et~al.}{2012}]{parsons2012}
Parsons A.~R.,  Pober J.~C.,  Aguirre J.~E.,  Carilli C.~L.,  Jacobs D.~C.,
  Moore D.~F.,  2012, \apj, 756, 165

\bibitem[\protect\citeauthoryear{Peebles}{Peebles}{1980}]{peebles1980}
Peebles P. J.~E.,  1980, The large-scale structure of the universe.
Princeton university press

\bibitem[\protect\citeauthoryear{Pober et~al.,}{Pober et~al.}{2013}]{pober2013}
Pober J.~C.,  et~al., 2013, \aj, 145, 65

\bibitem[\protect\citeauthoryear{Prasad \& Subrahmanya}{Prasad \&
  Subrahmanya}{2011a}]{prasad2011a}
Prasad P.,  Subrahmanya C.,  2011a, arXiv:1102.0148

\bibitem[\protect\citeauthoryear{Prasad \& Subrahmanya}{Prasad \&
  Subrahmanya}{2011b}]{prasad2011b}
Prasad P.,  Subrahmanya C.,  2011b, Experimental Astronomy, 31, 1

\bibitem[\protect\citeauthoryear{Prochaska \& Wolfe}{Prochaska \&
  Wolfe}{2009}]{prochaska2009}
Prochaska J.~X.,  Wolfe A.~M.,  2009, \apj, 696, 1543

\bibitem[\protect\citeauthoryear{Rao, Turnshek  \& Nestor}{Rao
  et~al.}{2006}]{rao2006}
Rao S.~M.,  Turnshek D.~A.,   Nestor D.~B.,  2006, \apj, 636, 610

\bibitem[\protect\citeauthoryear{Rhee, Zwaan, Briggs, Chengalur, Lah, Oosterloo
   \& van~der Hulst}{Rhee et~al.}{2013}]{rhee2013}
Rhee J.,  Zwaan M.~A.,  Briggs F.~H.,  Chengalur J.~N.,  Lah P.,  Oosterloo T.,
    van~der Hulst T.,  2013, \mnras, p. stt1481

\bibitem[\protect\citeauthoryear{Rhee, Lah, Chengalur, Briggs  \& Colless}{Rhee
  et~al.}{2016}]{rhee2016}
Rhee J.,  Lah P.,  Chengalur J.~N.,  Briggs F.~H.,   Colless M.,  2016, \mnras,
  p. stw1097

\bibitem[\protect\citeauthoryear{Santos et~al.,}{Santos
  et~al.}{2015}]{santos2015}
Santos M.~G.,  et~al., 2015, arXiv:1501.03989

\bibitem[\protect\citeauthoryear{Sarkar \& Datta}{Sarkar \&
  Datta}{2015}]{guhasarkar-datta2015}
Sarkar T.~G.,  Datta K.~K.,  2015, \jcap, 2015, 001

\bibitem[\protect\citeauthoryear{{Sarkar}, {Bharadwaj}  \&
  {Anathpindika}}{{Sarkar} et~al.}{2016}]{bharadwaj-sarkar2016}
{Sarkar} D.,  {Bharadwaj} S.,   {Anathpindika} S.,  2016, \mn@doi [\mnras]
  {10.1093/mnras/stw1111}, \href
  {http://adsabs.harvard.edu/abs/2016MNRAS.460.4310S} {460, 4310}

\bibitem[\protect\citeauthoryear{{Sarkar}, {Bharadwaj}  \& {Ali}}{{Sarkar}
  et~al.}{2017}]{sarkar-fisher2017}
{Sarkar} A.~K.,  {Bharadwaj} S.,   {Ali} S.~S.,  2017, \mn@doi [\japa]
  {10.1007/s12036-017-9432-2}, \href
  {http://adsabs.harvard.edu/abs/2017JApA...38...14S} {38, 14}

\bibitem[\protect\citeauthoryear{Sarma, Joshi, Bagri  \& Ananthakrishnan}{Sarma
  et~al.}{1975}]{sarma1975}
Sarma N.,  Joshi M.,  Bagri D.,   Ananthakrishnan S.,  1975, IETE Journal of
  Research, 21, 110

\bibitem[\protect\citeauthoryear{Seo, Dodelson, Marriner, Mcginnis, Stebbins,
  Stoughton  \& Vallinotto}{Seo et~al.}{2010}]{seo2010}
Seo H.-J.,  Dodelson S.,  Marriner J.,  Mcginnis D.,  Stebbins A.,  Stoughton
  C.,   Vallinotto A.,  2010, \apj, 721, 164

\bibitem[\protect\citeauthoryear{Storrie-Lombardi, McMahon  \&
  Irwin}{Storrie-Lombardi et~al.}{1996}]{storrie1996}
Storrie-Lombardi L.,  McMahon R.,   Irwin M.,  1996, \mnras, 283, L79

\bibitem[\protect\citeauthoryear{{Subrahmanya}, {Prasad}, {Girish},
  {Somashekar}, {Manoharan}  \& {Mittal}}{{Subrahmanya}
  et~al.}{2017a}]{subrahmanya2017b}
{Subrahmanya} C.~R.,  {Prasad} P.,  {Girish} B.~S.,  {Somashekar} R.,
  {Manoharan} P.~K.,   {Mittal} A.~K.,  2017a, \mn@doi [\japa]
  {10.1007/s12036-017-9434-0}, \href
  {http://adsabs.harvard.edu/abs/2017JApA...38...11S} {38, 11}

\bibitem[\protect\citeauthoryear{Subrahmanya, Manoharan  \&
  Chengalur}{Subrahmanya et~al.}{2017b}]{subrahmanya2017a}
Subrahmanya C.,  Manoharan P.,   Chengalur J.~N.,  2017b, \japa, 38, 10

\bibitem[\protect\citeauthoryear{Swarup et~al.,}{Swarup
  et~al.}{1971}]{swarup1971}
Swarup G.,  et~al., 1971, Nature Physical Science, 230, 185

\bibitem[\protect\citeauthoryear{Swarup, Ananthakrishnan, Kapahi, Rao,
  Subrahmanya  \& Kulkarni}{Swarup et~al.}{1991}]{Swarup1991}
Swarup G.,  Ananthakrishnan S.,  Kapahi V.,  Rao A.,  Subrahmanya C.,
  Kulkarni V.,  1991, Current Science, 60, 95

\bibitem[\protect\citeauthoryear{Switzer et~al.,}{Switzer
  et~al.}{2013}]{switzer2013}
Switzer E.,  et~al., 2013, \mnras: Letters, 434, L46

\bibitem[\protect\citeauthoryear{Thompson, Moran  \& Swenson~Jr}{Thompson
  et~al.}{2008}]{thompson2008}
Thompson A.~R.,  Moran J.~M.,   Swenson~Jr G.~W.,  2008, Interferometry and
  synthesis in radio astronomy.
John Wiley \& Sons

\bibitem[\protect\citeauthoryear{Tingay et~al.,}{Tingay et~al.}{2013}]{mwa2013}
Tingay S.,  et~al., 2013, \pasa, 30

\bibitem[\protect\citeauthoryear{Villaescusa-Navarro, Bull  \&
  Viel}{Villaescusa-Navarro et~al.}{2015a}]{navarro2015}
Villaescusa-Navarro F.,  Bull P.,   Viel M.,  2015a, \apj, 814, 146

\bibitem[\protect\citeauthoryear{Villaescusa-Navarro, Viel, Alonso, Datta, Bull
   et~al.}{Villaescusa-Navarro et~al.}{2015b}]{navarro-viel2015}
Villaescusa-Navarro F.,  Viel M.,  Alonso D.,  Datta K.~K.,  Bull P.,   et~al.,
  2015b, \jcap, 2015, 034

\bibitem[\protect\citeauthoryear{Villaescusa-Navarro, Alonso  \&
  Viel}{Villaescusa-Navarro et~al.}{2016}]{navarro-viel2016}
Villaescusa-Navarro F.,  Alonso D.,   Viel M.,  2016, arXiv:1609.00019

\bibitem[\protect\citeauthoryear{Visbal, Loeb  \& Wyithe}{Visbal
  et~al.}{2009}]{visbal2009}
Visbal E.,  Loeb A.,   Wyithe S.,  2009, \jcap, 2009, 030

\bibitem[\protect\citeauthoryear{Wolfe, Gawiser  \& Prochaska}{Wolfe
  et~al.}{2005}]{wolfe2005}
Wolfe A.~M.,  Gawiser E.,   Prochaska J.~X.,  2005, \araa, 43, 861

\bibitem[\protect\citeauthoryear{Wyithe \& Loeb}{Wyithe \&
  Loeb}{2008}]{wyithe-loeb2008}
Wyithe J. S.~B.,  Loeb A.,  2008, \mnras, 383, 606

\bibitem[\protect\citeauthoryear{Wyithe, Loeb  \& Geil}{Wyithe
  et~al.}{2008}]{wyithe2008}
Wyithe J. S.~B.,  Loeb A.,   Geil P.~M.,  2008, \mnras, 383, 1195

\bibitem[\protect\citeauthoryear{Yatawatta et~al.,}{Yatawatta
  et~al.}{2013}]{lofar2013}
Yatawatta S.,  et~al., 2013, \aap, 550, A136

\bibitem[\protect\citeauthoryear{Zafar, P{\'e}roux, Popping, Milliard,
  Deharveng  \& Frank}{Zafar et~al.}{2013}]{zafar2013}
Zafar T.,  P{\'e}roux C.,  Popping A.,  Milliard B.,  Deharveng J.-M.,   Frank
  S.,  2013, \aap, 556, A141

\bibitem[\protect\citeauthoryear{Zhang, Zuo, Ansari, Chen, Li, Wu, Campagne  \&
  Magneville}{Zhang et~al.}{2016}]{zhang2016}
Zhang J.,  Zuo S.,  Ansari R.,  Chen X.,  Li Y.,  Wu F.,  Campagne J.-E.,
  Magneville C.,  2016, arXiv:1606.03830

\bibitem[\protect\citeauthoryear{Zwaan, Meyer, Staveley-Smith  \&
  Webster}{Zwaan et~al.}{2005}]{zwaan2005}
Zwaan M.,  Meyer M.,  Staveley-Smith L.,   Webster R.,  2005, \mnras: Letters,
  359, L30

\makeatother
\end{thebibliography}

\bsp	
\label{lastpage}
\end{document}